# Self-Organization In Stellar Evolution: Size-Complexity Rule

Travis Herman Butler
Biological and Physical Sciences Department
Assumption University
Worcester, MA, USA 01609

Georgi Yordanov Georgiev
Physics Department
Assumption University and Worcester Polytechnic Institute
Worcester, MA, USA 01609
ggeorgie@assumption.edu, ggeorgiev@wpi.edu,
georgi@alumni.tufts.edu

Abstract

Complexity Theory is highly interdisciplinary, therefore any regularities must hold on all levels of organization, independent on the nature of the system. An open question in science is how complex systems self-organize to produce emergent structures and properties, a branch of non-equilibrium thermodynamics. It has long been known that there is a quantity-quality transition in natural systems. This is to say that the properties of a system depend on its size. More recently, this has been termed the size-complexity rule, which means that to increase their size, systems must increase their complexity, and that to increase their complexity they must grow in size. This rule goes under different names in different disciplines and systems of different nature, such as the area-speciation rule, economies of scale, scaling relations (allometric) in biology and for cities, and many others. We apply the size-complexity rule to stars to compare them with other complex systems in order to find universal patterns of self-organization independent of the substrate. Here, as a measure of complexity of a star, we are using the degree of grouping of nucleons into atoms, which reduces nucleon entropy, increases the variety of elements, and changes the structure of the star. As seen in our previous work, complexity, using action efficiency, is in power law proportionality of all other characteristics of a complex system, including its size. Here we find that, as for the other systems studied, the complexity of stars is in a power law proportionality with their size - the bigger a system is, the higher its level of complexity is - despite differing explosion energies and initial metallicities from simulations and data, which confirms the size-complexity rule and our model.

Introduction

This paper studies the phenomenon of self-organization in nature, its mechanisms and connections to other characteristics of complex systems. The subject of quantity-quality transition dates back to Aristotle with the famous quote that "the whole is something besides the parts" (Aristotle, 2006), but most of the Ancient Greek ideas they credited to have imported from Ancient Egypt, Babylon, Sumer and possibly other ancient civilizations such as ancient India and China. Therefore, the origin of those ideas is possibly from prerecorded times, which proves that it has been tested again and again at different levels of understanding. It was given its name by Hegel who implies a causal relation between the quality and quantity of self-organizing systems, or a quantity-quality transition (Hegel, 2014). More recently this has been termed a Size-Complexity rule (Bonner, 2004). We apply our previous model in which each characteristic is proportional to each other characteristic of the complex system as it grows, not just the size and complexity, following power law dependencies (G. Y. Georgiev, Chatterjee, & Iannacchione, 2017; G. Y. Georgiev et al., 2015). While there are so many principles and connections to discuss for different characteristics and different systems, which we will do in future papers, we will focus on only one - the size-complexity relationship - and its expression in nucleosynthesis during stellar evolution. Self-organization occurs only in systems far from thermodynamic equilibrium.

Complexity (quality), is an intensive property, which makes sense for a point or the parts of the system. Size (quantity), is an extensive property that describes the whole system. Size and complexity describe different aspects of self-organization, reinforcing each other in a positive feedback loop, as we see in our model (G. Y. Georgiev et al., 2015). Self-organizing systems cannot exist without size and complexity causing an increase each other, with some exceptions for simpler systems. This is further shown in other work where Eric Chaisson has correlated the free energy rate density (FERD) in complex systems to their level of complexity and evolutionary stage, another size-complexity rule, and used the term Cosmic Evolution to describe this process (Chaisson, 2002). The relevance of this connection to our work is that FERD is used to do the work to structure the system and build and maintain its complexity, an aspect which we will explore in future papers.

There are a variety of names for the size-complexity rule, as seen in the examples. The literature on size-structure relation is enormous, which confirms its proven and established rules and its universality. We only mention some examples. We just want to reiterate that the size-complexity rule is another term for quantity-quality transition, in evolving systems outside of thermodynamic equilibrium of physical, chemical, biological or social nature. Their difference stems from the fact that they were rediscovered in different areas of science independently. They all study the same universal phenomenon in complex systems. This shows that this rule is identical in all systems and is universal, something that is a centerpiece of our research - to search for universality of complexity across systems of different nature.

This research has broader connection to other areas of science. In this paper, we study the progress of self-organization in stars as a function of their size. We use the degree of grouping of nucleons into elements by nucleosynthesis (progress of nucleosynthesis and therefore self-organization) as a measure of complexity. We chose this measure because it is clear and unambiguous, metallicity data are widely available for stars and galaxies by observations and simulations, can be related to other measures in other systems, and can be readily calculated from data. We calculate this progress in stellar evolution for stars of different masses and initial metallicities and with different explosion energies. The purpose of this is to test our model (G. Y. Georgiev et al., 2017, 2015) and to compare it with the processes self-organization increase in other systems. We consider stars as complex systems, and one measure of their complexity is the degree to which they combine nucleons into heavier elements. The more of the heavier elements there are as a fraction of the mass of the star, the more advanced it is in its evolutionary stage and degree of complexity. There could be other measures, such as the differentiation of the internal structure of the star, but, they will be related to this one. We chose simulations by (Nomoto, Tominaga, Umeda, Kobayashi, & Maeda, 2006) of stars of different masses, explosion energies and metallicities at the end of their life, when they have exploded as supernovae. Those simulations were checked against observations of already exploded stars in the SAGA catalog where stellar composition can be measured by spectral analysis of their nebulae (Suda et al., 2008).

## Stellar Evolution Overview

The energy from the Big Bang self-organized under the influence of the strong, weak, and electromagnetic forces into matter, with 75% of which was Hydrogen and the rest was primarily Helium, with trace amounts of Lithium. Eventually, large amounts of hydrogen atoms coalesced and created strong gravity centers, which forced these atoms together in a dense space and heated them up. This fused many of the hydrogen atoms together, releasing vast amounts of energy. The heat force opposes the force of gravity to keep a star in equilibrium and continues to add nucleons to existing elements to form new elements through nucleosynthesis, the process of self-organization we are studying. Stellar nucleosynthesis begins after the gravitational collapse of a dense, molecular cloud into a protostar. The mass of the protostar determines if it will reach the temperatures necessary for nuclear fusion and star formation. Towards the end of a star's life, it inefficiently forms heavier elements beyond iron, which absorbs heat and energy instead of releasing it. Eventually, the star no longer has sufficient heat to oppose gravity, and it collapses in on itself and explodes (Thielemann, Diehl, Heger, Hirschi, & Liebendörfer, 2018). During this explosion, heavier elements, such as gold and platinum, are synthesized. Therefore, the complex systems that are being examined in this paper are stars synthesizing elements throughout the course of their lifetime, including when they explode in a supernova event. The elemental abundances can then be detected using spectrophotometry.

## Other Stellar Studies and Models

Other research has been done on stars to see how their nucleosynthesis and supernova event describes the chemical evolution of our galaxy using observational data of supernovae and metal-poor stars (Nomoto, Kobayashi, & Tominaga, 2013). Similar research has been done on how these yields (Nomoto et al., 1997, 2013, 2006) are affected by hydrodynamic effects during hypernova and supernova explosions (Nomoto & Suzuki, 2013). Moreover, studies have been done on nucleosynthesis to accurately understand the abundance pattern of Pop III stars leading to hypernovae (Nomoto, 2016). In addition, nucleosynthesis yields have been used to distinguish high-density Chandrasekhar-mass models and lower-density white dwarfs (Mori et al., 2018). These studies are important because they show patterns of chemical evolution and nucleosynthesis (Nomoto, 2016; Nomoto et al., 2013). The degree of grouping of nucleons as a definition for complexity is inversely proportional to nucleon entropy which decreases during nucleosynthesis (de Avellar, de Souza, & Horvath, 2016).

Later simulations have looked at and utilized the heavy-flavor neutrinos emitted from proton stars, like SN1987A, for triggered parameter explosions to more accurately report chemical evolution and iron group nucleosynthesis yields of in proto-neutron stars, taking into consideration the electron fraction of the ejecta (Curtis et al., 2018; Sinha et al., 2017). Another simulation has been done to look at the stellar yields of the first supernovae in stars of 12 to 140 solar masses and how rotation affects the nucleosynthesis yields (Takahashi, Umeda, & Yoshida, 2014). Simulations on nucleosynthetic yield for asymptotic giant branch, white dwarf, and core collapsing stars have been performed (Pignatari et al., 2016; Ritter et al., 2018). We find that the model of Nomoto 2006 is the best for studying nucleosynthesis because it shows the abundances of isotopes of various metallicities and solar massed stars as suggested by other reports (Wanajo, Nomoto, Janka, Kitaura, & Müller, 2009). More recent studies of nucleosynthesis have not shown the chemical abundances of isotopes when initial metallicities and explosion energies vary. These results are useful in studying the size-complexity rule amongst stars of varying size, initial metallicities, and explosion energies. Many of the recent papers above have discussed abundance levels in different simulated stars. However, none have reported their star's yields as thoroughly as Nomoto.

Hypothesis

Earlier research has shown that the size-complexity rule is valid in a variety of non-equilibrium thermodynamic systems, as evident in biological cells that form spherical structures (Amado, Batista, & Campos, 2018; Bell & Mooers, 1997). Research has been conducted to investigate the size-complexity in the life cycle (Bonner, 1995, 2015). In general, these studies demonstrate that there are two ways to measure complexity increase. One is the differentiation as a function of size, and the other is the structure formation (integration) in the larger system as a result of this differentiation. For example, the number of different cells in an organism is one measure, but, it is correlated to the structure formation in organs and the overall functioning of the organism. There are rare exceptions, such as of slime mold which can consist of one cell, but the vast majority of organisms follow the general rule (Bonner, 2004; McCarthy & Enquist, 2005).

In stellar evolution, nucleosynthesis produces atoms that are different than those that initially existed, and in general, the larger a star is, the more variety of atoms it can produce. This, in turn, works in a manner analogous to complexity in biological systems, with larger systems having more internal differentiation, which for stars with a large variety of elements determines layered regions inside the star of different density, temperature, kinds of atoms, and nuclear reactions occurring there. The differentiation leads to change in the global overall structure in the system, as observed in stars, organisms, cities, economies, etc. We argue that the existing physics laws, with extension and modification, are sufficient to describe self-organization in all those different systems (Walker, 2019).

Thus, our hypothesis, based on our previous model and data, (G. Y. Georgiev et al., 2017, 2015) is that larger stars will have greater progress of nucleosynthesis, complexity of element structure, at the end of their lives. This is because higher mass stars are much hotter, and the gravitational force is greater, causing greater pressure and density of matter at the core, which allows them to fuse more nucleons in their shorter life.

This has analogs in other complex systems, based on the size-complexity rule, which states that a system's complexity is contingent on its size. As we noted in our previous publications on Core Processing Units (CPUs) evolution, the level of organizational complexity and size were found to be locked in a positive feedback loop, and consequently, both increase exponentially through time, as a power law function of each other (G. Y. Georgiev et al., 2017, 2015). Our research aims to determine whether there is a similar trend in stars undergoing nucleosynthesis or not.

Theory

Model and Overview of Previous Work

Here we include for reference the basic model which studies how two of the characteristics (observables, descriptors, measures, properties): size and complexity depend on each other and what time behaviour can be predicted in general for complex systems (G. Y. Georgiev et al., 2015). In this previously published paper,(G. Y. Georgiev et al., 2015), we labeled $\alpha$ as a measure of complexity, or level of organization, and Q is the measure of the size of the system. The measure of complexity $\alpha$ is the average action efficiency per one event in the system based on the Principle of Least Action (for any event in the universe, action tends towards a least value) which in this model drives self-organization forward and Q is the total amount of action of the system(G. Georgiev & Georgiev, 2002; G. Y. Georgiev & Chatterjee, 2016; G. Y. Georgiev et al., 2017, 2016, 2015). In this paper, the analogs to those measures are progress and mass. To achieve higher complexity, the system needs to have larger size. This model explores a positive feedback loop (reinforcing) between the quality and quantity, or size and complexity(G. Y. Georgiev et al., 2015). It is supported by the observations about the size-complexity rule(Bell & Mooers, 1997; Bonner, 2004; Carneiro, 1967) and size-efficiency rules (Bejan, Lorente, Yilbas, & Sahin, 2011; Kleiber et al., 1932; West, Brown, & Enquist, 1999). An increase in an extensive property of the size of a system, drives it further away from equilibrium. This allows more work to be done to organize, leading to higher levels of organization reflected in its action efficiency per one event, $\alpha$ as the numerical measure for organization. Here we only present the results that are relevant to this application of the above-developed model. The solutions of the model are an exponential growth of quantity and quality in time, and a proportionality between them, which obeys a power law equation. In this paper, we study only the power law proportionality of quality from quantity, as the data presented do not have time resolution with the lifetime of the star, and the size of stars are fixed by the primordial gas cloud. Therefore, the size cannot grow in response to the increase of the complexity of the star, which is one exception of our model, because stars are simpler systems than biological and social. Nevertheless, stellar systems obey the size-complexity rule, as their structure is dependent on their size, as it will be seen in the results of this paper.

The exponential growth in time for the model is verified by data for CPUs (G. Y. Georgiev et al., 2015). Both quantities, $\alpha$ and $Q$, fit very well with exponentials on log-linear plots. Eliminating time, a solution of the exponential equations is the power law relation (equation 1).

$$\alpha = \eta \cdot Q^\gamma. \qquad (1)$$

Where $\gamma$ and $\eta$ are constants coming from the solutions of the exponential equations (G. Y. Georgiev et al., 2015).

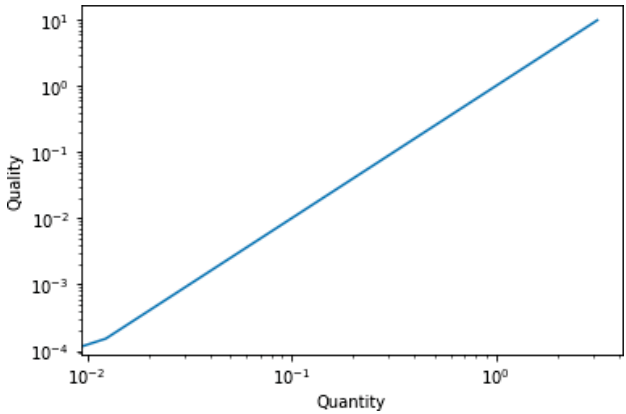

*Figure 1*. Numerical solution of quality (organization, complexity) vs. quantity (size of the system) from the system of differential equations of our model, in arbitrary units.

Figure 1 displays the numerical solution of the power law proportionality prediction of our model, shown in equation 1 (G. Y. Georgiev et al., 2015), for how the quality of stars in terms of progress of nucleosynthesis and internal organization and nucleon entropy reduction will depend on size, on a log-log plot. Here, for brevity, we just show one dependency from all possible extensive measures of size, (such as total free energy rate density, total power, total number of nucleons, etc.), namely mass. Therefore, in this paper, we test whether stars obey the prediction that their complexity at the end of their life is dependent on their mass, as a power law. Our full model predicts that the progress of nucleosynthesis will be exponential during the stellar lifetime, but, there are no data yet to test this prediction. However, since it is connected to the power law solutions, confirming them will strengthen the model and, therefore, the exponential growth prediction.

The goal of this continuous study is to find universality - mechanisms that are unchanged across the broadest range of systems of different nature, having explanatory power for all of them. The size-complexity rule is only one of the dependencies in the model of self-organization (G. Y. Georgiev et al., 2017). Each characteristic of a complex system depends on all others, not just the size and complexity, by a power law. In the general model, quantity accumulation leads to quality increase, but also, quality increase leads to quantity accumulation, which is the case in more advanced systems, such as biological and social, but not in Stellar Evolution.

Data and Methods

Data and Previous Simulations

The data represent the nucleosynthesis yields as a function of initial metallicity and stellar mass from nucleosynthesis yields of core collapse supernovae and galactic chemical evolution. We use it to check whether the progress of nucleosynthesis obeys a power law as a function of size, as shown in the previous model (G. Y. Georgiev et al., 2015). These yields are based on the new developments in the observational and theoretical studies of supernovae and extremely metal-poor stars in the halo, which have provided excellent opportunities to test the explosion models and their nucleosynthesis (Nomoto et al., 2006). In this paper, the initial metallicities of the stars studied are 0.000, 0.001, 0.004, and 0.02. Their masses are 13, 15, 18, 20, 25, 30, and 40 solar masses, when only the metallicity varies. Additionally, we also study when these stars have varying explosion energies. The masses for these are 20, 25, 30, and 40 solar masses, with a metallicity 0.000, 0.001, 0.004, and 0.02. This simulation is of large stars that explode as supernovae, and their composition can be compared with the measurements of the composition of already exploded stars in the Stellar Abundances for Galactic Archaeology (SAGA) Database(Suda et al., 2008). Only stars of mass larger than eight solar masses explode as supernovae, that is why there are no data for less massive stars.

Much research about stellar systems has been devoted to collecting the elemental abundances of stars from nearby galaxies. Several reports (Frebel et al., 2005; Umeda & Nomoto, 2003) have discussed that abundances in HE0107-5240 and other extremely metal poor stars are in good accord with nucleosynthesis that happens in 20 to 130 solar mass stars. Elemental abundances for smaller stars, such as red giants, were reported in Omega Centauri (Johnson & Pilachowski, 2010). Other research (Wanajo et al., 2009) has studied the yield and nucleosynthesis of unstable elements and reported their abundance for ST and FP3 model stars.

Further simulations (Tominaga, Iwamoto, & Nomoto,

2014) presented Pop III SN models, in which nucleosynthesis yields individually reproduce the abundance patterns of 48 metal-poor stars. Observations of abundances found in extremely metal-poor stars, HE 1300+0157, have also been done (Frebel et al., 2007). Another study (Prantzos, Abia, Limongi, Chieffi, & Cristallo, 2018) shows evolution of abundance of elements from Hydrogen to Uranium occurring in the Milky Way halo through a chemical evolution model of metallicity dependent isotopic yields from large stars. Some chemical abundances of extremely metal-poor stars from Pop III stars have been shown to describe the nature of first generation stars formed after the Big Bang (Nomoto, Tominaga, Umeda, & Kobayashi, 2005; Steigman, 2007).

Abundance information is critical to our research because, using stellar calculations (Nomoto et al., 2006) based on known abundances (Suda et al., 2008), we determined the progress of nucleosynthesis of elements at the end of the stars' life. Other investigators' search for abundances of exploded and simulated stars allows us to apply our model to their findings and see how efficient nucleosynthesis is in both massive and small-scale stars.

## Methods

The stellar yields of various isotopes, ranging from $^1H$ to $^{71}Ga$, were taken from 13, 15, 18, 20, 25, 30, and 40 solar mass stars with varying metallicities: 0, 0.001, 0.004, and 0.02. These yields of each isotope were given in solar masses from the SAGA Database (Suda et al., 2008) and (Nomoto et al., 2006). From the raw data of (Nomoto et al., 2006), the number of solar masses of each isotope from $^1H$ to $^{71}Ga$ and elements heavier than $^{71}Ga$ present within each star at the end of its life was first converted to the total number of nucleons present within each star as a measure of its size. In our calculations of progress of nucleosynthesis, we excluded the Hydrogen and Helium isotopes that existed before the star was formed when calculating the progress of nucleosynthesis. To determine the amount of Helium produced by the stars, we used Equation 2:

$$He_{nuc}[M_\odot] = \left(\frac{\psi}{M_*} - \frac{25.2}{100}\right) \cdot M_* \qquad (2)$$

(All of the symbols are listed in the footnote.) [1]

The total amount of Helium in each star was summed and divided by the total number of solar masses to obtain its fraction from the stars' mass. Then the fraction of Helium originally present immediately after the Big Bang was subtracted from it to find the fraction of the star made of synthesized Helium by the star in solar masses (Wagoner, Fowler, & Hoyle, 1967). This number is reported as $He_{nuc}$. For this fraction, we used a mass number of four, because that isotope is the majority of Helium present in the star.

To more accurately calculate the progress of nucleosynthesis, we also included the elements heavier than $^{71}Ga$ even though Nomoto (Nomoto et al., 2006) excluded them, because those would have only been measured in trace amounts. When the estimates of all of those isotopes are added together, they form a significant fraction of the masses of the stars. Because information for these elements was not provided, the mass number of these is assumed to be 140, because that is approximately the mean mass number between Gallium and Uranium in the periodic system. These elements are referred to as "$^{140}\chi$". The total mass of $^{140}\chi$ was calculated by subtracting from the total initial mass of the star the $^1H$ to $^{71}Ga$, the equivalent mass fraction of the explosion energy, and $M_{cut}$, all in the same units of solar masses. To find the mass of the explosion energy, the following equation was used Einstein (1905):

$$E = M \cdot c^2 \qquad (3)$$

where $E$ is energy in joules, $M$ is the mass in kilograms, and $c$ is the speed of light in vacuum. This mass was then converted to solar masses by dividing it by the mass of the sun in kilograms.

Additionally, we do not integrate the $M_{cut}$ mass value of Nomoto et al. (2006) into our calculation for progress of nucleosynthesis because of the lack of information about its exact composition. This is due to the fact that the material in the nebula contains information about all elements present in the star, but not the ones remaining in the central remnant object. The figures and trend line calculations were plotted and fitted using Python Spyder version 3.6.1.

## Initial Metallicity

In the following calculations, we took into account the initial metallicity of the star to exclude those heavier than

---

[1] Symbols: $He_{nuc}[M_\odot]$ is the mass of the helium nucleosynthesized in the star in solar masses. $n$ distinguishes the isotopes of all elements. $\theta_n$ is the total number of nucleons for each isotope, $n$, within the star. $\theta_i$ is the total number of nucleons present in each star. $M_\odot$ is a unit for the number of solar masses. $M_\odot[kg]$ is the mass of the Sun in kilograms. $M_{is}$ is the mass of each isotope present in the star in solar masses. $M_{cut}$ is the total mass of the remnant after the explosion of the supernova at the center of the nebula in solar masses. $N_A$ is Avagadro's number. $\psi$ is the combined solar masses of Helium-3 and Helium-4. $M_*$ is the mass of a star in solar masses. $^{140}\chi$ represents the mass of the elements more advanced than $^{71}Ga$ in solar masses. $^{100}\rho$ is the mass of the heavier than helium elements that were made by a previous generation star (based on its metallicity). $\varrho_n$ is the degree of grouping of nucleons for a selected individual isotope. $A$ is the mass number of a selected isotope. $\varrho_{sum}$ is grouping of all nucleons in each star. $P$ is the progress of nucleosynthesis, the grouping of all nucleons in each star, normalized by its size (total number of nucleons).

Helium elements that were synthesized by a previous star. Because these specific elements are not listed, we assume that they have a mass number of 100 and factor them into these calculations as "$^{100}\rho$." 100 is the approximate mean mass number of all of the elements in the periodic system, excluding hydrogen and the pre-synthesized helium, taking into account their relative abundance. To calculate the number of solar masses of $^{100}\rho$, the initial metallicity of each star as a fraction of the stellar mass was multiplied by its $M_*$, the star's mass in solar masses.

To find the progress of nucleosynthesis, we first calculated the number of nucleons of each isotope present in the star at the end of its life using Equation 4 for each studied star.

$$\beta_n = M_\odot[kg] \cdot M_{is}[M_\odot]_n \cdot 10^3[\frac{g}{kg}] \cdot N_A \quad (4)$$

The number of nucleons, $\beta$, for each isotope, $n$, for the stars with metallicities: 0, 0.001, 0.004, and 0.02, are shown in Tables 3, 4, 5, and 6, respectively.

We calculated from its mass the total number of nucleons present in each star, $\beta_i$, which is a measure of its size, with the equation 5:

$$\beta_i = M_\odot[kg] \cdot M_*[M_\odot] \cdot 10^3[\frac{g}{kg}] \cdot N_A \quad (5)$$

The total number of nucleons, $\beta_i$, for stars of different masses are shown in Table 7. We then multiplied the number of nucleons of each isotope for synthesized Helium, and for all isotopes from $^6Li$ to $^{71}Ga$, and $^{140}\chi$ by its mass number, which is the number of nucleons of each individual isotope, to obtain $\varrho_n$ as a measure of the degree of grouping of the nucleons, using equation 6:

$$\varrho_n = \beta_n \cdot A \quad (6)$$

where $A$ is the mass number of the isotope. To get $\varrho_{sum}$ (equation 7), we took the $\varrho_n$ of each isotope of synthesized Helium, $\varrho_{He}$, the sum of all isotopes from $^6Li$ to $^{71}Ga$, $\sum_{n=6}^{71} \varrho_n$, the number of nucleons of $^{140}\chi$, $\varrho_\chi$, and added them together. From this number we subtracted the number of nucleons of $^{100}\rho$, $\varrho_\rho$ to avoid including isotopes that were not synthesized by the star. The mass of $M_{cut}$, explosion energy, and the pre-existing Hydrogen and Helium in the star do not participate in this calculation.

$$\varrho_{sum} = \varrho_{He} + \sum_{n=6}^{71} \varrho_n + \varrho_\chi - \varrho_\rho \quad (7)$$

The level of progress (complexity) for how far stars went in grouping nucleons together into heavier isotopes over their lifetime was determined. The more connected the nucleons are, the more advanced the nucleosynthesis is in terms of the degree of complexity and progress of filling the periodic system by that star. The progress of nucleosynthesis for each star, $P$, is then determined by dividing $\varrho_{sum}$ by the total number of nucleons in stars 13, 15, 18, 20, 25, 30, and 40 $M_\odot$ (eq. 8):

$$P = \frac{\varrho_{sum}}{\beta_i} \quad (8)$$

The progress of nucleosynthesis was determined for each star when the metallicity fraction equals 0, 0.001, 0.004, and 0.02 and reported in Table 9.

Figure 2 shows the progress of nucleosynthesis versus the initial number of $M_*$ with initial metallicities equal to 0, 0.001, 0.004, and 0.02.

Varying Explosion Energies

The stellar yields of various isotopes, ranging from $^1H$ to $^{71}Ga$, were taken from 20, 25, 30, and 40 solar massed stars with varying explosion energies and metallicities of 0, 0.001, 0.004, and 0.02. Stars with 20 and 25 solar masses have 10 E of explosion energy, where in all cases, $E \sim 1x10^{51} ergs$. 30 solar massed stars have 20E of explosion energy. Stars of 40 solar masses have explosion energies of 30E. These yields of each isotope were given in solar masses from the SAGA Database (Suda et al., 2008) and (Nomoto et al., 2006). The method used for determining the $P$ of varying metallicities stars is the same for when the metallicity and explosion energy vary. Table 10 shows the $\varrho_{sum}$ of each star when the explosion energy varies and the metallicity equals 0, 0.001, 0.004 and 0.02 . The progress of nucleosynthesis is reported in Table 11.

Results

In this section, we present a study of stars with varying initial metallicities, without taking into account the explosion energies. Tables 3, 4, 5, and 6 show the calculated number of nucleons for each isotope present in stars of various masses (Nomoto et al., 2006), when the initial metallicities equal 0, 0.001, 0.004, and 0.02 respectively. Table 7 shows the total number of nucleons of each star, assumed constant throughout its life.

Table 8 shows the $\varrho_{sum}$ of stars of each mass when their metallicity varies. Table 9 shows the progress of nucleosynthesis when the initial metallicity is 0, 0.001, 0.004, and 0.02 respectively. Figure 2 shows the progress of nucleosynthesis versus the number of solar masses for stars of each mass and metallicity on a Log/Log scale at the end of the stars' life. This shows that the progress of nucleosynthesis follows a power law in stars even when their initial metallicities vary. The advance in the progress of nucleosynthesis for each metallicity is similar as the trend lines show. The significance that they follow a power law is that it matches our previous model of interdependence of the characteristics of complex systems and the power law proportionality between them, which is confirmed empirically by all data for size-complexity rules by Bonner, Carneiro, and others. (Bonner, 2004; Carneiro, 1967; G. Y. Georgiev et al., 2017, 2015). We use this study on one hand to additionally test empirically our previously published model, and on the other hand to apply that model to learn more about stellar evolution and to be able to extrapolate it to predict how larger or smaller stars will evolve based on what we have learned from stars between 13 and 40 solar masses. We are trying to gain the largest generality of the model and also use what we learn from stars to apply it to other complex systems as listed elsewhere in the text. The predictions of our previous model are for power law relationships, and the data for stellar evolution from simulations done by Nomoto support those predictions, which further strengthens the model. This increases our level of confidence in the model and in its applicability for other systems.

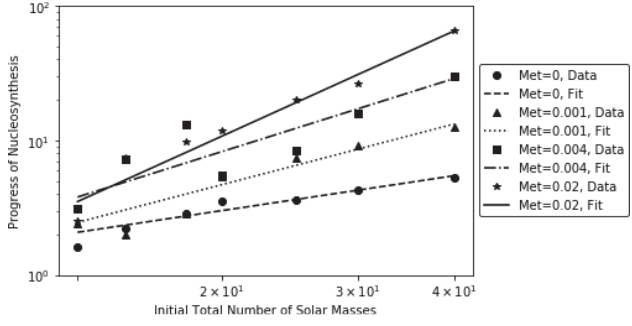

*Figure 2*. Progress of nucleosynthesis vs. mass on a Log/Log scale, indicating a power law proportionality, confirming our model.

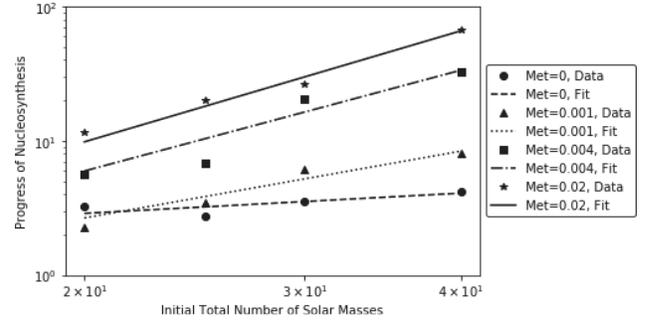

*Figure 3*. Progress of nucleosynthesis vs. mass on a Log/Log scale when the explosion energies vary.

Table 1 shows the coefficients, represented by *c*, and the powers, represented by *y*, for the fits on Figure 2, where the progress of nucleosynthesis is a function of the initial number of nucleons. These equations follow the format:

$$P = c \cdot M_*[M_\odot]^y \qquad (9)$$

Within the data of this simulation, there is an additional trend that the higher the initial metallicity, the greater the increase in progress as a function of mass. As it can be seen in table 1, the powers of the fits increase from 0.86 to 2.6 for stars from 0 initial metallicity to stars of 0.02 initial metallicity.

Varying Explosion Energy, Varying Initial Metallicity

Table 11 shows the progress of nucleosynthesis within each star when explosion energies vary and the initial metallicity is 0, 0.001, 0.004, and 0.02, respectively. Figure 3 shows the progress of nucleosynthesis versus the mass of stars in solar masses for each of the given metallicities at their explosion energies on a Log/Log scale at the end of each star's life. This shows that the progress of nucleosynthesis follows a power law in stars even when their initial metallicities and explosion energies vary, as the trend lines show.

Table 1
*Coefficients and Powers for the Progress vs. Solar Mass Trend Line of Figure 2*

| Metallicity | Coefficient | Power |
|---|---|---|
| 0 | $2.28 \times 10^{-1}$ | 0.86 |
| 0.001 | $5.74 \times 10^{-2}$ | 1.47 |
| 0.004 | $3.71 \times 10^{-2}$ | 1.80 |
| 0.02 | $4.48 \times 10^{-3}$ | 2.60 |

Table 2 shows the coefficients and the powers of the equations of Figure 3 where the progress of nucleosynthesis is function of the stellar mass. These equations follow the format of equation 9.

We observe that the progress of nucleosynthesis is increasing with the size of the star as measured by its mass, which confirms the size-complexity rule, as reported by other authors (Bonner, 2004; Carneiro, 1967). Within the data of this simulation, there is an additional trend that the higher the initial metallicity, the greater the increase in progress as a function of mass. As it can be seen in table 2, the powers of the fits increase from 0.5 to 2.75 for stars from 0 initial metallicity to stars of 0.02 initial metallicity.

Table 2
*Coefficients and Powers for the Progress vs. Solar Mass Trend Line of Figure 3 when Explosion Energies Vary*

| Metallicity | Coefficient | Power |
|---|---|---|
| 0 | $6.42 \times 10^{-1}$ | 0.50 |
| 0.001 | $1.88 \times 10^{-2}$ | 1.65 |
| 0.004 | $3.37 \times 10^{-3}$ | 2.50 |
| 0.02 | $2.61 \times 10^{-3}$ | 2.75 |

There are a total of only four points in the graphs describing trends in stars with varying explosion energy, compared to the seven points in graphs where explosion energy is not taken into account. This is because (Nomoto et al., 2006) exclude 13 to 18 solar massed stars from their yields tables when the explosion energy varies.

## Discussion

Figure 2 shows that despite the variation in the initial metallicity of the star, its progress of nucleosynthesis at the end of its life increases when the size of the star increases. This means that the system's size is a determinant of its level of self-organization. This can be compared to the size-complexity rule in all other self-organized systems of any nature - the number of ant castes as a function of the size of an ant colony, the number of different types of cells as a function of the size of an organism (Bonner, 2004), the number of organizational traits in societies as a function of the size of those societies (Carneiro, 1967), the species diversity as a function of the area size of the ecosystem, the number of species as a function of the size of a population and the size of the area that they occupy (Maurer, 1996), the number of occupations in a human society as a function of the number of people (Bonner, 2004), the GDP per person in cities as a function of city size (West, 2017), the action efficiency of CPUs (G. Y. Georgiev et al., 2015) and almost any other example for any other system. The trends presented on figure 2 confirm the size-complexity rule, and match well with our model of self-organization, as first applied for CPUs (G. Y. Georgiev et al., 2017, 2015). We challenge the reader to find more examples or systems where this proportionality is not valid, as any exceptions will be very informative and improve the model.

Figure 3 suggests that the progress of nucleosynthesis follows a power law regardless of the variation of the initial metallicities and explosion energies of stars. This means that the level of complexity is more advanced in higher mass than in lower mass stars at the end of their life, even when explosion energies are taken into account, similar to other complex systems as mentioned above (Bonner, 2004; G. Y. Georgiev et al., 2015).

There is an additional trend that the higher the initial metallicity, the greater the increase in progress as a function of mass, with or without taking into account the explosion energies, as it can be seen in tables 1 and 2. In our model (G. Y. Georgiev et al., 2015), self-organization progresses exponentially during the lifetime of the system. Therefore, if the stellar system starts at a higher level of organization, then, it is further along its exponential trend, and, continuing that trend from a higher level, reaches much higher values of complexity within the same lifetime of the stars. This is seen visually, in the increase of the slopes of figures 2 and 3 with increased initial metallicity, and in the increase of the power of the corresponding fit equation in the tables 1 and 2. Those numbers can be used to calculate the exponential trends within the lifetime of stars, a subject of future work.

To reiterate, we found that the progress is in a power law dependence on the mass of the star, even when the initial metallicity and explosion energies are different. This compares to our previous studies of CPU evolution, where all of their characteristics were demonstrated to be power laws of each other (G. Y. Georgiev et al., 2017, 2015). This shows that the larger the star is, the greater its progress of nucleosynthesis, as a measure for its level of complexity, similar to previous studies of size-complexity rules (Bonner, 2004). By calculating the stellar lifetime, using the mass of each star, it is also possible to find the average rate of nucleosynthesis over their lifetime, including the final explosion and the action efficiency of nucleosynthesis as a function of the mass of stars. We will show this in follow up papers. The goal is to show that this power law dependence and the exponential increase during the lifetime are universal features of all complex systems, independent of their nature, be they physical, chemical, biological, technological, social, etc., and to apply it in our future work to as many other systems as possible to look for confirmations of this model and for possible exceptions.

Because in our model, the power law relations between all characteristics of complex systems are always obeyed, this may be one criteria that can be used to recognize which systems are complex. The rest, which are simpler, do not self-organize, and we cannot use any of their characteristics, to predict any other. There is a continuum of less complex and more complex systems, as some of the less complex obey certain aspects of the model, but not others. Notably, stars cannot grow in size in response of the increase of their complexity, as their mass is fixed by the primordial gas cloud. Other intermediate systems that obey only some aspects of the model are physical systems, such as Benard Cells and Vortexes, and chemical systems, such as Belousov–Zhabotinsky (BZ) reactions.

## Conclusions

Our conclusions are not only about stellar evolution, but, complex systems and complexification in general. We check our model (G. Y. Georgiev et al., 2015) with a specific system - stars - to understand the process of self-organization better. Stellar evolution obeys the size-complexity rule (Bonner, 2004), because the progress of nucleosythesis as measured by the degree of grouping of nucleons is in a power law of stellar mass. Progress of nucleosynthesis confirms one aspect of our model, namely the quantity-quality transition (G. Y. Georgiev et al., 2017, 2015) and the other aspects of the mutual dependence of their characteristics, such as rate of nucleosynthesis, number of nuclear additions, Free Energy Rate Density, etc. are to be explored in future papers. This gives us a new understanding of stars as complex systems. It has predictive power for progress of nucleosynthesis and the changes in the other stellar characteristics dependent on the masses of stars outside of the range included in this study, and within the lifetime of each star, which are new hypotheses to be tested. This strengthens our model and allows us to make predictions about the general behavior of other complex systems.

In this paper, we aim for generality through a very specific example. The succession of nucleosynthesis is pre-programmed in natural laws and is repeated in all stars of similar mass, as is the self-organization of all other systems. We cannot predict which two atoms will interact at any given moment, but, we can predict the global statistical properties of the star, as we can predict that the number of species will increase if we increase the size of an ecosystem, or the number of different occupations will increase if a city grows in size, without being able to predict any specifics. It is in the true sense evolution - "unfolding" of something that can be predicted and is prescribed upfront towards a "future state", in the case of stars, of heavier nuclei, in the case of ecosystems - of a larger variety of species, or in the case of cities - of increased variety of occupations and organization traits, etc. There is a small room for chance or contingency on external fluctuations, within the limits of stability of the system.

Ultimately, the existence of a complex system causes its self-organization and its self-organization causes its existence. In the example that we investigated, stellar self-organization is the natural byproduct of the star's existence, the nucleons group and increase the complexity of the star while providing energy for its existence. On the other hand, the existence of a star is a byproduct of its self-organization - nucleosynthesis is what creates the energy to provide the outward pressure to balance gravity and maintain its existence while on the main sequence. Size-complexity examples, such as those that operate in biology, are that if cells do not differentiate, organisms cease to grow and if they do not grow, their cells cannot differentiate (Bonner, 2004). This is true statistically for most of the organisms, as there are small exceptions of it, such as for slime mold. Furthermore, Carneiro shows that if new organizational units do not appear in expanding societies, they split into smaller ones, and if they do not grow in size they cannot acquire new organizational traits. There are animal species, such as alligators and turtles, and cities, such as Rome or Venice, that appear to have existed without changing, but this is only true over relatively short time-scales. If we extend the time-scale to the length of the entire evolution of that system - billions of years in biology and thousands of years in the human civilization, we will see that they always change, although at varying rates. We may even go to such lengths, as to say that civilizations that stopped self-organizing (improving, evolving), were overtaken by the ones that did, and therefore many of those ceased to exist. We can make a prediction that those trends will continue in the future as long as there are non-equilibrium thermodynamic systems of any kind, and that this happens everywhere in the Universe.

In order to show the applicability of first principles, and of the Principle of Least Action and action efficiency importance in this research, as in our model (G. Y. Georgiev et al., 2017, 2015), we are planning to look for the time dependence of the rate of nucleosynthesis during the lifetime in each individual star. Based on our model, the prediction is that complexity (amount of grouping of nucleons, the progress of nucleosynthesis) increases exponentially during the star's lifetime. Another prediction is that action efficiency of nucleosynthesis will increase following a power law with the mass of stars. Whether those predictions are confirmed or rejected will either strengthen or modify our model. Another prediction based on our model is that when a gas cloud collapses into a star, there is a positive feedback between the gravity and the amount of matter in the star, therefore initially, the size of the star will increase exponentially, and then as the matter in the gas cloud is exhausted, it will reach saturation and stop growing, forming a logistic curve.

Stellar evolution is naturally connected to galactic evolution, as the metallicity of galaxies and their structure formation depends on the processes of nucleosynthesis in stars. In our future work we will explore whether galactic evolution also obeys the size-complexity rule and all other proportionality relations between the characteristics of complex systems and their exponential increase in time, to test and develop further our model.

## Acknowledgements

We would like to thank the Assumption University Honor's Program and the Department of Biological and Physical Sciences for financial support, as well as Assumption students Thanh Vu, Simon Trcka and Lamberto Qako for help with data collection. We would also like to thank Assumption students Nicole Bramlitt and Rebecca Gilchrist for long discussions and for proofreading the paper, and to Prof. Steven Theroux, Prof. Michael Levin, Clement Vidal and Claudio Flores Martinez for valuable feedback.

Table 3
*Calculated Number of Nucleons of Each Isotope when the Metallicity Equals 0*

| $M_*(M_\odot)$ | 13 | 15 | 18 | 20 | 25 | 30 | 40 |
|---|---|---|---|---|---|---|---|
| $^1H$ | $7.89\times10^{57}$ | $9.08\times10^{57}$ | $1.01\times10^{58}$ | $1.05\times10^{58}$ | $1.27\times10^{58}$ | $1.4\times10^{58}$ | $1.68\times10^{58}$ |
| $^2H$ | $1.78\times10^{41}$ | $2.02\times10^{41}$ | $1.53\times10^{41}$ | $1.04\times10^{41}$ | $2.42\times10^{41}$ | $1.61\times10^{41}$ | $4.14\times10^{41}$ |
| $^3He$ | $4.93\times10^{52}$ | $4.9\times10^{52}$ | $3.99\times10^{52}$ | $5.7\times10^{52}$ | $2.53\times10^{53}$ | $2.47\times10^{53}$ | $3.07\times10^{52}$ |
| $^4He$ | $4.8\times10^{57}$ | $5.27\times10^{57}$ | $6.49\times10^{57}$ | $7.11\times10^{57}$ | $9.62\times10^{57}$ | $1.14\times10^{58}$ | $1.43\times10^{58}$ |
| $He_{nuc}$ | $8.79\times10^{56}$ | $7.43\times10^{56}$ | $1.06\times10^{57}$ | $1.08\times10^{57}$ | $2.07\times10^{57}$ | $2.35\times10^{57}$ | $2.18\times10^{57}$ |
| $^6Li$ | $4.37\times10^{34}$ | $1.33\times10^{35}$ | $5.23\times10^{34}$ | $4.37\times10^{36}$ | $3.22\times10^{36}$ | $1.35\times10^{35}$ | $9.03\times10^{35}$ |
| $^7Li$ | $2.6\times10^{47}$ | $3.52\times10^{47}$ | $8.79\times10^{46}$ | $3.34\times10^{47}$ | $6.8\times10^{48}$ | $2.83\times10^{49}$ | $4.5\times10^{46}$ |
| $^9Be$ | $2.12\times10^{37}$ | $3.86\times10^{35}$ | $1.26\times10^{35}$ | $5.38\times10^{34}$ | $1.49\times10^{40}$ | $1.51\times10^{37}$ | $6.38\times10^{37}$ |
| $^{10}B$ | $3.5\times10^{36}$ | $9.94\times10^{37}$ | $4.7\times10^{36}$ | $1.88\times10^{38}$ | $3.44\times10^{39}$ | $6.2\times10^{37}$ | $2.84\times10^{40}$ |
| $^{11}B$ | $3.52\times10^{41}$ | $3.95\times10^{41}$ | $8.55\times10^{41}$ | $7.83\times10^{40}$ | $1.13\times10^{42}$ | $3.92\times10^{42}$ | $3.68\times10^{43}$ |
| $^{12}C$ | $8.88\times10^{55}$ | $2.06\times10^{56}$ | $2.61\times10^{56}$ | $2.53\times10^{56}$ | $3.52\times10^{56}$ | $4.04\times10^{56}$ | $5.14\times10^{56}$ |
| $^{13}C$ | $1\times10^{50}$ | $7.44\times10^{49}$ | $3.15\times10^{48}$ | $1.37\times10^{49}$ | $1.76\times10^{49}$ | $1.22\times10^{49}$ | $3.84\times10^{48}$ |
| $^{14}N$ | $2.19\times10^{54}$ | $2.23\times10^{54}$ | $2.26\times10^{53}$ | $6.49\times10^{52}$ | $7.08\times10^{53}$ | $1.96\times10^{51}$ | $7.05\times10^{50}$ |
| $^{15}N$ | $7.64\times10^{49}$ | $8.22\times10^{49}$ | $2.87\times10^{49}$ | $1.35\times10^{49}$ | $1.4\times10^{50}$ | $2\times10^{49}$ | $7.53\times10^{50}$ |
| $^{16}O$ | $5.39\times10^{56}$ | $9.26\times10^{56}$ | $1.65\times10^{57}$ | $2.53\times10^{57}$ | $3.34\times10^{57}$ | $5.76\times10^{57}$ | $1\times10^{58}$ |
| $^{17}O$ | $2.02\times10^{51}$ | $1.88\times10^{51}$ | $3.34\times10^{50}$ | $8.18\times10^{49}$ | $1.78\times10^{51}$ | $2.25\times10^{49}$ | $1.7\times10^{48}$ |
| $^{18}O$ | $6.94\times10^{49}$ | $5.86\times10^{51}$ | $5.55\times10^{51}$ | $3.02\times10^{49}$ | $8.08\times10^{50}$ | $2.47\times10^{48}$ | $2.55\times10^{50}$ |
| $^{19}F$ | $1.4\times10^{47}$ | $2.36\times10^{48}$ | $9.47\times10^{48}$ | $1.94\times10^{48}$ | $2.05\times10^{48}$ | $1.07\times10^{48}$ | $2.85\times10^{47}$ |
| $^{20}Ne$ | $1.83\times10^{55}$ | $3.92\times10^{56}$ | $5.92\times10^{56}$ | $1.09\times10^{57}$ | $6.38\times10^{56}$ | $1.02\times10^{57}$ | $3.68\times10^{56}$ |
| $^{21}Ne$ | $6.49\times10^{50}$ | $4.5\times10^{52}$ | $1.09\times10^{53}$ | $5.15\times10^{52}$ | $1.59\times10^{52}$ | $6.6\times10^{52}$ | $1.29\times10^{52}$ |
| $^{22}Ne$ | $2.37\times10^{50}$ | $1.93\times10^{52}$ | $3.08\times10^{52}$ | $8.29\times10^{52}$ | $2.42\times10^{52}$ | $1.03\times10^{53}$ | $8.08\times10^{51}$ |
| $^{23}Na$ | $1.72\times10^{53}$ | $2.93\times10^{54}$ | $2.49\times10^{54}$ | $3.47\times10^{54}$ | $1.23\times10^{54}$ | $1.7\times10^{54}$ | $2.2\times10^{53}$ |
| $^{24}Mg$ | $1.03\times10^{56}$ | $8.17\times10^{55}$ | $1.88\times10^{56}$ | $1.8\times10^{56}$ | $1.44\times10^{56}$ | $2.71\times10^{56}$ | $5.73\times10^{56}$ |
| $^{25}Mg$ | $1.87\times10^{53}$ | $3.57\times10^{53}$ | $6.98\times10^{53}$ | $1.39\times10^{53}$ | $4.76\times10^{52}$ | $2.92\times10^{53}$ | $5.13\times10^{53}$ |
| $^{26}Mg$ | $8.47\times10^{52}$ | $4.77\times10^{53}$ | $1.05\times10^{54}$ | $2.85\times10^{53}$ | $6\times10^{52}$ | $1.55\times10^{53}$ | $1.5\times10^{53}$ |
| $^{26}Al$ | $1.19\times10^{51}$ | $1.33\times10^{51}$ | $3.99\times10^{51}$ | $5.96\times10^{50}$ | $8.98\times10^{50}$ | $3.5\times10^{51}$ | $1.66\times10^{51}$ |
| $^{27}Al$ | $4.53\times10^{54}$ | $1.64\times10^{54}$ | $3.76\times10^{54}$ | $1.64\times10^{54}$ | $9.68\times10^{53}$ | $3.15\times10^{54}$ | $1.76\times10^{55}$ |
| $^{28}Si$ | $9.63\times10^{55}$ | $8.77\times10^{55}$ | $1.39\times10^{56}$ | $1.19\times10^{56}$ | $4.2\times10^{56}$ | $2.97\times10^{56}$ | $1.22\times10^{57}$ |
| $^{29}Si$ | $8.98\times10^{53}$ | $2.86\times10^{53}$ | $5.29\times10^{53}$ | $2.18\times10^{53}$ | $3.25\times10^{53}$ | $7.04\times10^{53}$ | $3.11\times10^{54}$ |
| $^{30}Si$ | $1.7\times10^{54}$ | $1.78\times10^{53}$ | $4.13\times10^{53}$ | $1.32\times10^{53}$ | $9.03\times10^{52}$ | $3.05\times10^{53}$ | $4.86\times10^{54}$ |
| $^{31}P$ | $5.85\times10^{53}$ | $6.74\times10^{52}$ | $1.58\times10^{53}$ | $9.59\times10^{52}$ | $1.01\times10^{53}$ | $1.4\times10^{53}$ | $1.92\times10^{54}$ |
| $^{32}S$ | $2.84\times10^{55}$ | $3.83\times10^{55}$ | $4.87\times10^{55}$ | $6.36\times10^{55}$ | $2.22\times10^{56}$ | $1.39\times10^{56}$ | $4.47\times10^{56}$ |
| $^{33}S$ | $1.08\times10^{53}$ | $9.04\times10^{52}$ | $1.23\times10^{53}$ | $2.37\times10^{53}$ | $3.28\times10^{53}$ | $1.98\times10^{53}$ | $9.7\times10^{53}$ |
| $^{34}S$ | $3.34\times10^{53}$ | $2.42\times10^{53}$ | $3.41\times10^{53}$ | $5.87\times10^{53}$ | $5.08\times10^{53}$ | $1.01\times10^{53}$ | $1.9\times10^{54}$ |
| $^{36}S$ | $1.77\times10^{49}$ | $1.71\times10^{48}$ | $6.4\times10^{48}$ | $3.07\times10^{48}$ | $4.08\times10^{47}$ | $8.43\times10^{47}$ | $3.82\times10^{49}$ |
| $^{35}Cl$ | $6.56\times10^{52}$ | $1.75\times10^{52}$ | $3.16\times10^{52}$ | $8.24\times10^{52}$ | $6.49\times10^{52}$ | $2.75\times10^{52}$ | $2.59\times10^{53}$ |
| $^{37}Cl$ | $3.64\times10^{51}$ | $6.98\times10^{51}$ | $1.09\times10^{52}$ | $4.62\times10^{52}$ | $7.32\times10^{52}$ | $1.84\times10^{52}$ | $1.16\times10^{53}$ |
| $^{36}Ar$ | $3.88\times10^{54}$ | $6.32\times10^{54}$ | $6.79\times10^{54}$ | $1.16\times10^{55}$ | $3.71\times10^{55}$ | $2.35\times10^{55}$ | $5.83\times10^{55}$ |
| $^{38}Ar$ | $6.26\times10^{52}$ | $7.46\times10^{52}$ | $2.04\times10^{53}$ | $4.62\times10^{53}$ | $4.3\times10^{53}$ | $4.17\times10^{52}$ | $1.33\times10^{54}$ |
| $^{40}Ar$ | $9.59\times10^{46}$ | $2.13\times10^{46}$ | $4.74\times10^{46}$ | $1.27\times10^{47}$ | $2.12\times10^{46}$ | $5.58\times10^{45}$ | $1.55\times10^{47}$ |
| $^{39}K$ | $6.01\times10^{51}$ | $9.41\times10^{51}$ | $2.31\times10^{52}$ | $5.27\times10^{52}$ | $7.55\times10^{52}$ | $1.56\times10^{52}$ | $1.41\times10^{53}$ |
| $^{40}K$ | $1.37\times10^{48}$ | $1.09\times10^{48}$ | $2.35\times10^{48}$ | $1.56\times10^{49}$ | $9.63\times10^{48}$ | $1.19\times10^{48}$ | $1.25\times10^{49}$ |
| $^{41}K$ | $4.24\times10^{50}$ | $9.64\times10^{50}$ | $2.06\times10^{51}$ | $1.13\times10^{52}$ | $2.49\times10^{52}$ | $4.06\times10^{51}$ | $3.35\times10^{52}$ |
| $^{40}Ca$ | $3.5\times10^{54}$ | $5.28\times10^{54}$ | $5.27\times10^{54}$ | $7.45\times10^{54}$ | $2.97\times10^{55}$ | $2.08\times10^{55}$ | $4.47\times10^{55}$ |
| $^{42}Ca$ | $1.17\times10^{51}$ | $1.47\times10^{51}$ | $4.34\times10^{51}$ | $1.53\times10^{52}$ | $8.94\times10^{51}$ | $1.03\times10^{51}$ | $2.6\times10^{52}$ |
| $^{43}Ca$ | $7.74\times10^{49}$ | $5.91\times10^{49}$ | $4.06\times10^{49}$ | $8.9\times10^{49}$ | $1.89\times10^{49}$ | $2.31\times10^{48}$ | $1.16\times10^{49}$ |
| $^{44}Ca$ | $2.01\times10^{52}$ | $2.65\times10^{52}$ | $1.74\times10^{52}$ | $1.74\times10^{52}$ | $1.17\times10^{52}$ | $6.52\times10^{51}$ | $1.04\times10^{52}$ |
| $^{46}Ca$ | $1.28\times10^{45}$ | $2.11\times10^{45}$ | $1.11\times10^{46}$ | $1.56\times10^{47}$ | $3.33\times10^{46}$ | $7.25\times10^{44}$ | $6.23\times10^{45}$ |
| $^{48}Ca$ | $1.86\times10^{40}$ | $5.04\times10^{43}$ | $4.93\times10^{41}$ | $5.27\times10^{41}$ | $1.51\times10^{46}$ | $4.25\times10^{41}$ | $1.56\times10^{40}$ |
| $^{45}Sc$ | $2.54\times10^{49}$ | $4.83\times10^{49}$ | $6.64\times10^{49}$ | $3.05\times10^{50}$ | $7.38\times10^{50}$ | $1.81\times10^{50}$ | $6.96\times10^{50}$ |
| $^{46}Ti$ | $7.52\times10^{51}$ | $3.25\times10^{51}$ | $4.86\times10^{51}$ | $7.16\times10^{51}$ | $3.81\times10^{51}$ | $6.32\times10^{50}$ | $1.29\times10^{52}$ |
| $^{47}Ti$ | $1.05\times10^{52}$ | $4.62\times10^{51}$ | $6.35\times10^{51}$ | $5.1\times10^{51}$ | $7.07\times10^{49}$ | $4.55\times10^{49}$ | $1.11\times10^{50}$ |
| $^{48}Ti$ | $7.59\times10^{52}$ | $1\times10^{53}$ | $9.16\times10^{52}$ | $1.06\times10^{53}$ | $1.86\times10^{53}$ | $2.16\times10^{53}$ | $2.93\times10^{53}$ |
| $^{49}Ti$ | $2.72\times10^{51}$ | $3.98\times10^{51}$ | $3.59\times10^{51}$ | $4.72\times10^{51}$ | $8.43\times10^{51}$ | $1.06\times10^{52}$ | $1.44\times10^{52}$ |
| $^{50}Ti$ | $1.41\times10^{45}$ | $1.14\times10^{45}$ | $2.06\times10^{45}$ | $2.19\times10^{45}$ | $2.68\times10^{45}$ | $5.5\times10^{43}$ | $1.13\times10^{46}$ |
| $^{50}V$ | $1.62\times10^{46}$ | $1.26\times10^{46}$ | $5.51\times10^{46}$ | $1.63\times10^{47}$ | $1.4\times10^{47}$ | $3.16\times10^{45}$ | $7.82\times10^{47}$ |
| $^{51}V$ | $1.98\times10^{52}$ | $1.29\times10^{52}$ | $1.49\times10^{52}$ | $1.32\times10^{52}$ | $1.08\times10^{52}$ | $1.27\times10^{52}$ | $2.11\times10^{52}$ |
| $^{50}Cr$ | $1.25\times10^{52}$ | $1.86\times10^{52}$ | $3.16\times10^{52}$ | $2.66\times10^{52}$ | $5.85\times10^{52}$ | $4.58\times10^{52}$ | $2.13\times10^{53}$ |
| $^{52}Cr$ | $1.05\times10^{54}$ | $1.31\times10^{54}$ | $1.35\times10^{54}$ | $1.63\times10^{54}$ | $3.32\times10^{54}$ | $3.74\times10^{54}$ | $4.73\times10^{54}$ |
| $^{53}Cr$ | $5.94\times10^{52}$ | $8.1\times10^{52}$ | $7.7\times10^{52}$ | $9.92\times10^{52}$ | $1.8\times10^{53}$ | $2.18\times10^{53}$ | $3.13\times10^{53}$ |
| $^{54}Cr$ | $2.81\times10^{47}$ | $4.95\times10^{47}$ | $3.89\times10^{48}$ | $3.81\times10^{48}$ | $1.12\times10^{49}$ | $2.06\times10^{47}$ | $1.03\times10^{50}$ |
| $^{55}Mn$ | $1.59\times10^{53}$ | $2.22\times10^{53}$ | $2.08\times10^{53}$ | $2.71\times10^{53}$ | $5.15\times10^{53}$ | $6.18\times10^{53}$ | $8.56\times10^{53}$ |
| $^{54}Fe$ | $8.73\times10^{53}$ | $1.49\times10^{54}$ | $1.68\times10^{54}$ | $1.7\times10^{54}$ | $3.96\times10^{54}$ | $4.9\times10^{54}$ | $1.11\times10^{55}$ |
| $^{56}Fe$ | $8.38\times10^{55}$ | $8.38\times10^{55}$ | $8.38\times10^{55}$ | $8.38\times10^{55}$ | $8.38\times10^{55}$ | $8.38\times10^{55}$ | $8.41\times10^{55}$ |
| $^{57}Fe$ | $1.19\times10^{54}$ | $1.37\times10^{54}$ | $1.05\times10^{54}$ | $1.04\times10^{54}$ | $5.59\times10^{53}$ | $5.79\times10^{53}$ | $6\times10^{53}$ |
| $^{58}Fe$ | $6.79\times10^{46}$ | $2.11\times10^{47}$ | $9.67\times10^{47}$ | $1.26\times10^{48}$ | $2.77\times10^{48}$ | $2.71\times10^{47}$ | $1.69\times10^{49}$ |
| $^{59}Co$ | $2.11\times10^{53}$ | $1.58\times10^{53}$ | $1.93\times10^{53}$ | $1.8\times10^{53}$ | $1.88\times10^{52}$ | $2.96\times10^{51}$ | $2.24\times10^{51}$ |
| $^{58}Ni$ | $4.61\times10^{53}$ | $4.96\times10^{53}$ | $4.59\times10^{53}$ | $4.53\times10^{53}$ | $3.51\times10^{53}$ | $4.58\times10^{53}$ | $6.88\times10^{53}$ |
| $^{60}Ni$ | $2.54\times10^{54}$ | $1.94\times10^{54}$ | $1.88\times10^{54}$ | $1.61\times10^{54}$ | $1.77\times10^{53}$ | $4.31\times10^{51}$ | $5.25\times10^{51}$ |
| $^{61}Ni$ | $4.32\times10^{52}$ | $3.76\times10^{52}$ | $2.54\times10^{52}$ | $2.22\times10^{52}$ | $6.91\times10^{50}$ | $7.23\times10^{48}$ | $1.58\times10^{48}$ |
| $^{62}Ni$ | $2.32\times10^{52}$ | $1.81\times10^{52}$ | $1.62\times10^{52}$ | $1.43\times10^{52}$ | $4.65\times10^{50}$ | $1.96\times10^{48}$ | $6.42\times10^{47}$ |
| $^{64}Ni$ | $4.59\times10^{42}$ | $1.19\times10^{44}$ | $1.31\times10^{43}$ | $7.31\times10^{43}$ | $3.92\times10^{45}$ | $3.9\times10^{43}$ | $9.23\times10^{42}$ |
| $^{63}Cu$ | $5.86\times10^{51}$ | $4.24\times10^{51}$ | $4.7\times10^{51}$ | $4.12\times10^{51}$ | $1.47\times10^{50}$ | $2.84\times10^{47}$ | $2.47\times10^{46}$ |
| $^{65}Cu$ | $2.56\times10^{50}$ | $2.9\times10^{50}$ | $1.95\times10^{50}$ | $1.7\times10^{50}$ | $9.33\times10^{48}$ | $1.23\times10^{45}$ | $1.56\times10^{44}$ |
| $^{64}Zn$ | $1.5\times10^{53}$ | $1.46\times10^{53}$ | $1.14\times10^{53}$ | $9.93\times10^{52}$ | $3.04\times10^{51}$ | $3.62\times10^{47}$ | $5.22\times10^{46}$ |
| $^{66}Zn$ | $8.19\times10^{50}$ | $1.28\times10^{51}$ | $6.05\times10^{50}$ | $5.08\times10^{50}$ | $1.9\times10^{49}$ | $4.23\times10^{45}$ | $6.16\times10^{44}$ |
| $^{67}Zn$ | $1.94\times10^{49}$ | $2.79\times10^{49}$ | $1.56\times10^{49}$ | $1.26\times10^{49}$ | $2.59\times10^{47}$ | $7.56\times10^{44}$ | $1.01\times10^{44}$ |
| $^{68}Zn$ | $3.52\times10^{49}$ | $3.88\times10^{49}$ | $4.82\times10^{49}$ | $4.11\times10^{49}$ | $9.64\times10^{47}$ | $1.56\times10^{45}$ | $3.14\times10^{44}$ |
| $^{70}Zn$ | $8.31\times10^{41}$ | $4.3\times10^{43}$ | $7.89\times10^{42}$ | $3.01\times10^{43}$ | $4.76\times10^{43}$ | $2.23\times10^{43}$ | $5.33\times10^{41}$ |
| $^{69}Ga$ | $9.35\times10^{48}$ | $6.73\times10^{48}$ | $7.31\times10^{48}$ | $5.91\times10^{48}$ | $1.05\times10^{47}$ | $6.06\times10^{44}$ | $5.1\times10^{42}$ |
| $^{71}Ga$ | $1.02\times10^{43}$ | $1.34\times10^{44}$ | $2.2\times10^{43}$ | $1.1\times10^{44}$ | $2.68\times10^{44}$ | $2.8\times10^{43}$ | $1.63\times10^{42}$ |
| $^{140}\chi$ | 0 | 0 | 0 | 0 | 0 | 0 | 0 |
| $^{100}\rho$ | 0 | 0 | 0 | 0 | 0 | 0 | 0 |

Table 4
*Number of Nucleons of Each Isotope when the Metallicity Equals .001*

| $M_*(M_\odot)$ | 13 | 15 | 18 | 20 | 25 | 30 | 40 |
|---|---|---|---|---|---|---|---|
| $^1H$ | $7.71\times10^{57}$ | $8.92\times10^{57}$ | $1.01\times10^{58}$ | $1.01\times10^{58}$ | $1.17\times10^{58}$ | $1.32\times10^{58}$ | $1.55\times10^{58}$ |
| $^2H$ | $8.61\times10^{42}$ | $4.05\times10^{45}$ | $3.68\times10^{42}$ | $5.21\times10^{41}$ | $5.34\times10^{41}$ | $6.02\times10^{41}$ | $8.42\times10^{43}$ |
| $^3He$ | $1.71\times10^{53}$ | $1.83\times10^{53}$ | $1.88\times10^{53}$ | $1.92\times10^{53}$ | $1.51\times10^{53}$ | $1.72\times10^{53}$ | $1.45\times10^{53}$ |
| $^4He$ | $4.62\times10^{57}$ | $6.18\times10^{57}$ | $7.83\times10^{57}$ | $7.11\times10^{57}$ | $8.35\times10^{57}$ | $1.00\times10^{58}$ | $1.31\times10^{58}$ |
| $He_{nuc}$ | $7.00\times10^{57}$ | $1.65\times10^{57}$ | $2.04\times10^{57}$ | $1.08\times10^{57}$ | $8.03\times10^{56}$ | $9.82\times10^{56}$ | $9.82\times10^{56}$ |
| $^6Li$ | $2.84\times10^{40}$ | $1.33\times10^{40}$ | $1.06\times10^{40}$ | $2.87\times10^{34}$ | $2.81\times10^{36}$ | $3.29\times10^{34}$ | $2.74\times10^{41}$ |
| $^7Li$ | $7.59\times10^{47}$ | $4.50\times10^{44}$ | $6.98\times10^{44}$ | $1.01\times10^{47}$ | $1.09\times10^{45}$ | $1.34\times10^{45}$ | $8.00\times10^{45}$ |
| $^9Be$ | $2.79\times10^{40}$ | $1.09\times10^{42}$ | $1.12\times10^{41}$ | $1.71\times10^{34}$ | $2.46\times10^{32}$ | $0.00\times10^{00}$ | $4.86\times10^{41}$ |
| $^{10}B$ | $1.38\times10^{46}$ | $7.20\times10^{45}$ | $7.87\times10^{45}$ | $2.84\times10^{45}$ | $9.10\times10^{45}$ | $2.69\times10^{45}$ | $2.41\times10^{45}$ |
| $^{11}B$ | $6.12\times10^{46}$ | $3.21\times10^{46}$ | $3.52\times10^{46}$ | $1.23\times10^{46}$ | $4.10\times10^{46}$ | $1.21\times10^{46}$ | $8.40\times10^{45}$ |
| $^{12}C$ | $1.28\times10^{56}$ | $1.02\times10^{56}$ | $1.55\times10^{56}$ | $1.53\times10^{56}$ | $2.58\times10^{56}$ | $1.45\times10^{56}$ | $8.83\times10^{55}$ |
| $^{13}C$ | $2.00\times10^{53}$ | $6.44\times10^{52}$ | $8.62\times10^{52}$ | $2.35\times10^{52}$ | $1.18\times10^{53}$ | $9.82\times10^{52}$ | $3.38\times10^{53}$ |
| $^{14}N$ | $1.09\times10^{55}$ | $4.29\times10^{54}$ | $5.35\times10^{54}$ | $1.55\times10^{55}$ | $1.10\times10^{55}$ | $7.41\times10^{54}$ | $1.04\times10^{55}$ |
| $^{15}N$ | $9.07\times10^{51}$ | $1.03\times10^{51}$ | $1.10\times10^{51}$ | $1.64\times10^{51}$ | $8.67\times10^{51}$ | $4.54\times10^{50}$ | $2.92\times10^{51}$ |
| $^{16}O$ | $6.04\times10^{56}$ | $3.52\times10^{56}$ | $5.05\times10^{56}$ | $2.61\times10^{57}$ | $4.58\times10^{57}$ | $6.38\times10^{57}$ | $1.00\times10^{58}$ |
| $^{17}O$ | $8.34\times10^{52}$ | $3.07\times10^{52}$ | $2.90\times10^{52}$ | $2.61\times10^{52}$ | $3.34\times10^{52}$ | $5.83\times10^{52}$ | $3.43\times10^{52}$ |
| $^{18}O$ | $2.17\times10^{54}$ | $4.38\times10^{53}$ | $3.67\times10^{53}$ | $9.69\times10^{51}$ | $8.44\times10^{52}$ | $3.22\times10^{52}$ | $3.15\times10^{52}$ |
| $^{19}F$ | $3.68\times10^{51}$ | $2.37\times10^{50}$ | $3.98\times10^{50}$ | $3.35\times10^{51}$ | $7.33\times10^{50}$ | $2.86\times10^{51}$ | $5.35\times10^{51}$ |
| $^{20}Ne$ | $7.91\times10^{55}$ | $2.28\times10^{56}$ | $2.12\times10^{56}$ | $7.51\times10^{56}$ | $1.46\times10^{57}$ | $1.74\times10^{57}$ | $3.44\times10^{56}$ |
| $^{21}Ne$ | $2.22\times10^{53}$ | $8.08\times10^{52}$ | $1.05\times10^{53}$ | $1.64\times10^{53}$ | $6.10\times10^{53}$ | $7.32\times10^{53}$ | $1.33\times10^{53}$ |
| $^{22}Ne$ | $1.59\times10^{54}$ | $3.28\times10^{53}$ | $5.69\times10^{53}$ | $1.40\times10^{54}$ | $1.76\times10^{54}$ | $1.87\times10^{54}$ | $1.05\times10^{54}$ |
| $^{23}Na$ | $6.48\times10^{53}$ | $2.35\times10^{54}$ | $2.50\times10^{54}$ | $2.17\times10^{54}$ | $9.69\times10^{54}$ | $8.22\times10^{54}$ | $1.08\times10^{54}$ |
| $^{24}Mg$ | $7.62\times10^{55}$ | $7.63\times10^{55}$ | $7.10\times10^{55}$ | $2.90\times10^{56}$ | $2.14\times10^{56}$ | $3.43\times10^{56}$ | $8.43\times10^{56}$ |
| $^{25}Mg$ | $1.68\times10^{54}$ | $1.05\times10^{54}$ | $1.13\times10^{54}$ | $2.87\times10^{54}$ | $2.08\times10^{54}$ | $4.25\times10^{54}$ | $2.65\times10^{54}$ |
| $^{26}Mg$ | $9.87\times10^{53}$ | $1.37\times10^{54}$ | $1.11\times10^{54}$ | $2.91\times10^{54}$ | $2.38\times10^{54}$ | $5.10\times10^{54}$ | $1.32\times10^{54}$ |
| $^{26}Al$ | $1.21\times10^{52}$ | $2.19\times10^{51}$ | $3.73\times10^{51}$ | $4.28\times10^{51}$ | $3.33\times10^{51}$ | $9.17\times10^{51}$ | $5.99\times10^{51}$ |
| $^{27}Al$ | $4.28\times10^{54}$ | $2.81\times10^{54}$ | $2.77\times10^{54}$ | $8.36\times10^{54}$ | $6.05\times10^{54}$ | $1.05\times10^{55}$ | $3.61\times10^{55}$ |
| $^{28}Si$ | $1.08\times10^{56}$ | $5.14\times10^{55}$ | $1.83\times10^{56}$ | $1.53\times10^{56}$ | $1.44\times10^{56}$ | $1.98\times10^{56}$ | $1.06\times10^{57}$ |
| $^{29}Si$ | $1.70\times10^{54}$ | $4.58\times10^{53}$ | $7.20\times10^{53}$ | $1.33\times10^{54}$ | $5.27\times10^{53}$ | $1.23\times10^{54}$ | $7.23\times10^{54}$ |
| $^{30}Si$ | $2.22\times10^{54}$ | $5.09\times10^{53}$ | $6.40\times10^{53}$ | $9.29\times10^{53}$ | $3.29\times10^{53}$ | $8.49\times10^{53}$ | $1.21\times10^{55}$ |
| $^{31}P$ | $6.36\times10^{53}$ | $9.73\times10^{52}$ | $2.24\times10^{53}$ | $2.36\times10^{53}$ | $1.28\times10^{53}$ | $2.44\times10^{53}$ | $4.17\times10^{54}$ |
| $^{32}S$ | $4.43\times10^{55}$ | $1.96\times10^{55}$ | $9.44\times10^{55}$ | $6.73\times10^{55}$ | $6.60\times10^{55}$ | $9.34\times10^{55}$ | $3.94\times10^{56}$ |
| $^{33}S$ | $2.35\times10^{53}$ | $5.70\times10^{52}$ | $3.61\times10^{53}$ | $1.19\times10^{53}$ | $9.73\times10^{52}$ | $1.66\times10^{53}$ | $1.00\times10^{54}$ |
| $^{34}S$ | $1.10\times10^{54}$ | $2.86\times10^{53}$ | $7.63\times10^{53}$ | $3.71\times10^{53}$ | $2.14\times10^{53}$ | $4.41\times10^{53}$ | $2.54\times10^{54}$ |
| $^{36}S$ | $8.97\times10^{50}$ | $1.86\times10^{50}$ | $3.83\times10^{50}$ | $4.85\times10^{50}$ | $7.82\times10^{50}$ | $1.08\times10^{51}$ | $6.66\times10^{50}$ |
| $^{35}Cl$ | $9.77\times10^{52}$ | $9.70\times10^{51}$ | $9.77\times10^{52}$ | $2.79\times10^{52}$ | $2.52\times10^{52}$ | $4.01\times10^{52}$ | $4.22\times10^{53}$ |
| $^{37}Cl$ | $1.72\times10^{52}$ | $3.15\times10^{51}$ | $4.95\times10^{52}$ | $1.22\times10^{52}$ | $1.55\times10^{52}$ | $2.30\times10^{52}$ | $7.56\times10^{52}$ |
| $^{36}Ar$ | $6.82\times10^{54}$ | $2.95\times10^{54}$ | $1.46\times10^{55}$ | $1.21\times10^{55}$ | $1.11\times10^{55}$ | $1.62\times10^{55}$ | $5.46\times10^{55}$ |
| $^{38}Ar$ | $3.21\times10^{53}$ | $2.71\times10^{52}$ | $4.93\times10^{53}$ | $1.28\times10^{53}$ | $9.88\times10^{52}$ | $1.95\times10^{53}$ | $1.00\times10^{54}$ |
| $^{40}Ar$ | $2.11\times10^{50}$ | $5.87\times10^{49}$ | $1.23\times10^{50}$ | $6.50\times10^{49}$ | $1.18\times10^{50}$ | $1.41\times10^{50}$ | $8.18\times10^{49}$ |
| $^{39}K$ | $3.01\times10^{52}$ | $4.97\times10^{51}$ | $6.82\times10^{52}$ | $1.66\times10^{52}$ | $1.82\times10^{52}$ | $2.84\times10^{52}$ | $1.33\times10^{53}$ |
| $^{40}K$ | $3.19\times10^{49}$ | $2.34\times10^{48}$ | $4.70\times10^{49}$ | $9.07\times10^{48}$ | $1.19\times10^{49}$ | $1.14\times10^{49}$ | $2.85\times10^{49}$ |
| $^{41}K$ | $2.89\times10^{51}$ | $3.75\times10^{50}$ | $1.31\times10^{52}$ | $2.48\times10^{51}$ | $2.66\times10^{51}$ | $4.77\times10^{51}$ | $1.82\times10^{52}$ |
| $^{40}Ca$ | $5.67\times10^{54}$ | $2.07\times10^{54}$ | $9.65\times10^{54}$ | $1.10\times10^{55}$ | $9.51\times10^{54}$ | $1.40\times10^{55}$ | $4.38\times10^{55}$ |
| $^{42}Ca$ | $7.93\times10^{51}$ | $4.85\times10^{50}$ | $1.62\times10^{52}$ | $3.34\times10^{51}$ | $2.98\times10^{51}$ | $5.56\times10^{51}$ | $2.74\times10^{52}$ |
| $^{43}Ca$ | $4.97\times10^{50}$ | $4.01\times10^{50}$ | $2.93\times10^{50}$ | $2.29\times10^{50}$ | $3.47\times10^{50}$ | $4.25\times10^{50}$ | $3.37\times10^{50}$ |
| $^{44}Ca$ | $2.97\times10^{52}$ | $2.48\times10^{52}$ | $2.24\times10^{52}$ | $5.85\times10^{51}$ | $9.53\times10^{51}$ | $7.71\times10^{51}$ | $1.32\times10^{52}$ |
| $^{46}Ca$ | $7.97\times10^{49}$ | $1.69\times10^{49}$ | $3.05\times10^{49}$ | $4.01\times10^{49}$ | $3.62\times10^{49}$ | $8.49\times10^{49}$ | $1.46\times10^{50}$ |
| $^{48}Ca$ | $3.78\times10^{50}$ | $1.16\times10^{50}$ | $1.46\times10^{50}$ | $1.50\times10^{50}$ | $1.88\times10^{50}$ | $2.23\times10^{50}$ | $2.60\times10^{50}$ |
| $^{45}Sc$ | $3.02\times10^{50}$ | $6.08\times10^{49}$ | $4.25\times10^{50}$ | $2.59\times10^{50}$ | $3.88\times10^{50}$ | $5.05\times10^{50}$ | $1.19\times10^{51}$ |
| $^{46}Ti$ | $4.77\times10^{51}$ | $1.49\times10^{52}$ | $6.12\times10^{51}$ | $1.63\times10^{51}$ | $1.57\times10^{51}$ | $2.58\times10^{51}$ | $1.20\times10^{52}$ |
| $^{47}Ti$ | $3.05\times10^{51}$ | $1.53\times10^{52}$ | $1.80\times10^{51}$ | $5.15\times10^{50}$ | $4.02\times10^{50}$ | $5.39\times10^{50}$ | $7.61\times10^{50}$ |
| $^{48}Ti$ | $1.16\times10^{53}$ | $6.22\times10^{52}$ | $1.20\times10^{53}$ | $1.64\times10^{53}$ | $1.40\times10^{53}$ | $2.00\times10^{53}$ | $3.03\times10^{53}$ |
| $^{49}Ti$ | $4.61\times10^{51}$ | $2.32\times10^{51}$ | $4.95\times10^{51}$ | $8.12\times10^{51}$ | $6.80\times10^{51}$ | $1.01\times10^{52}$ | $1.68\times10^{52}$ |
| $^{50}Ti$ | $7.57\times10^{50}$ | $1.94\times10^{50}$ | $2.23\times10^{50}$ | $6.05\times10^{50}$ | $1.02\times10^{51}$ | $1.37\times10^{51}$ | $1.39\times10^{51}$ |
| $^{50}V$ | $1.32\times10^{49}$ | $1.34\times10^{48}$ | $2.17\times10^{48}$ | $1.02\times10^{49}$ | $1.57\times10^{49}$ | $2.47\times10^{49}$ | $1.11\times10^{50}$ |
| $^{51}V$ | $1.07\times10^{52}$ | $1.94\times10^{52}$ | $9.06\times10^{51}$ | $9.86\times10^{51}$ | $9.01\times10^{51}$ | $1.19\times10^{52}$ | $2.32\times10^{52}$ |
| $^{50}Cr$ | $2.52\times10^{52}$ | $1.59\times10^{52}$ | $3.58\times10^{52}$ | $2.84\times10^{52}$ | $3.29\times10^{52}$ | $4.20\times10^{52}$ | $1.62\times10^{53}$ |
| $^{52}Cr$ | $1.44\times10^{54}$ | $3.63\times10^{53}$ | $1.82\times10^{54}$ | $2.66\times10^{54}$ | $2.61\times10^{54}$ | $3.50\times10^{54}$ | $4.72\times10^{54}$ |
| $^{53}Cr$ | $8.90\times10^{52}$ | $3.38\times10^{52}$ | $1.09\times10^{53}$ | $1.64\times10^{53}$ | $1.33\times10^{53}$ | $2.12\times10^{53}$ | $3.07\times10^{53}$ |
| $^{54}Cr$ | $2.24\times10^{51}$ | $5.75\times10^{50}$ | $7.14\times10^{50}$ | $1.69\times10^{51}$ | $2.61\times10^{51}$ | $3.28\times10^{51}$ | $3.40\times10^{51}$ |
| $^{55}Mn$ | $2.72\times10^{53}$ | $9.59\times10^{52}$ | $3.37\times10^{53}$ | $4.55\times10^{53}$ | $3.58\times10^{53}$ | $6.00\times10^{53}$ | $9.41\times10^{53}$ |
| $^{54}Fe$ | $1.83\times10^{54}$ | $8.47\times10^{53}$ | $2.47\times10^{54}$ | $2.84\times10^{54}$ | $2.69\times10^{54}$ | $3.96\times10^{54}$ | $1.25\times10^{55}$ |
| $^{56}Fe$ | $8.70\times10^{55}$ | $8.48\times10^{55}$ | $8.52\times10^{55}$ | $8.49\times10^{55}$ | $8.52\times10^{55}$ | $8.53\times10^{55}$ | $8.56\times10^{55}$ |
| $^{57}Fe$ | $1.69\times10^{54}$ | $2.10\times10^{54}$ | $1.52\times10^{54}$ | $8.64\times10^{53}$ | $6.35\times10^{53}$ | $6.98\times10^{53}$ | $6.89\times10^{53}$ |
| $^{58}Fe$ | $6.80\times10^{52}$ | $1.57\times10^{52}$ | $1.86\times10^{52}$ | $5.98\times10^{52}$ | $9.32\times10^{52}$ | $1.11\times10^{53}$ | $1.22\times10^{53}$ |
| $^{59}Co$ | $1.07\times10^{53}$ | $2.78\times10^{53}$ | $7.55\times10^{52}$ | $7.44\times10^{52}$ | $4.10\times10^{52}$ | $6.46\times10^{52}$ | $5.29\times10^{52}$ |
| $^{58}Ni$ | $6.10\times10^{53}$ | $1.09\times10^{54}$ | $2.11\times10^{54}$ | $4.24\times10^{53}$ | $3.01\times10^{53}$ | $4.54\times10^{53}$ | $9.34\times10^{53}$ |
| $^{60}Ni$ | $1.94\times10^{54}$ | $3.08\times10^{54}$ | $1.40\times10^{54}$ | $3.53\times10^{53}$ | $7.87\times10^{52}$ | $2.32\times10^{53}$ | $1.21\times10^{53}$ |
| $^{61}Ni$ | $6.05\times10^{52}$ | $8.29\times10^{52}$ | $5.55\times10^{52}$ | $1.46\times10^{52}$ | $1.88\times10^{52}$ | $2.52\times10^{52}$ | $2.47\times10^{52}$ |
| $^{62}Ni$ | $1.21\times10^{53}$ | $2.24\times10^{53}$ | $3.56\times10^{53}$ | $3.03\times10^{52}$ | $4.44\times10^{52}$ | $6.17\times10^{52}$ | $9.10\times10^{52}$ |
| $^{64}Ni$ | $1.94\times10^{52}$ | $2.19\times10^{51}$ | $2.75\times10^{51}$ | $2.84\times10^{52}$ | $5.17\times10^{52}$ | $7.74\times10^{52}$ | $9.38\times10^{52}$ |
| $^{63}Cu$ | $9.64\times10^{51}$ | $6.55\times10^{51}$ | $3.33\times10^{51}$ | $1.15\times10^{52}$ | $1.72\times10^{52}$ | $2.36\times10^{52}$ | $1.68\times10^{52}$ |
| $^{65}Cu$ | $5.81\times10^{51}$ | $1.06\times10^{51}$ | $1.37\times10^{51}$ | $9.64\times10^{51}$ | $1.78\times10^{52}$ | $2.67\times10^{52}$ | $3.47\times10^{52}$ |
| $^{64}Zn$ | $8.56\times10^{52}$ | $7.79\times10^{52}$ | $6.37\times10^{52}$ | $1.40\times10^{52}$ | $1.08\times10^{52}$ | $1.93\times10^{52}$ | $1.55\times10^{52}$ |
| $^{66}Zn$ | $1.18\times10^{52}$ | $4.74\times10^{51}$ | $9.26\times10^{51}$ | $1.34\times10^{52}$ | $2.31\times10^{52}$ | $3.70\times10^{52}$ | $5.73\times10^{52}$ |
| $^{67}Zn$ | $1.25\times10^{51}$ | $1.69\times10^{50}$ | $2.24\times10^{50}$ | $2.44\times10^{51}$ | $4.62\times10^{51}$ | $7.26\times10^{51}$ | $5.01\times10^{51}$ |
| $^{68}Zn$ | $8.23\times10^{51}$ | $6.97\times10^{50}$ | $9.41\times10^{50}$ | $1.18\times10^{52}$ | $2.23\times10^{52}$ | $3.77\times10^{52}$ | $5.55\times10^{52}$ |
| $^{70}Zn$ | $2.74\times10^{50}$ | $1.26\times10^{49}$ | $2.10\times10^{49}$ | $3.43\times10^{49}$ | $2.81\times10^{49}$ | $8.83\times10^{49}$ | $1.66\times10^{50}$ |
| $^{69}Ga$ | $8.52\times10^{50}$ | $8.90\times10^{49}$ | $1.35\times10^{50}$ | $1.49\times10^{51}$ | $2.79\times10^{51}$ | $4.47\times10^{51}$ | $5.67\times10^{51}$ |
| $^{71}Ga$ | $7.59\times10^{50}$ | $6.14\times10^{49}$ | $9.22\times10^{49}$ | $1.31\times10^{51}$ | $2.41\times10^{51}$ | $4.60\times10^{51}$ | $6.53\times10^{51}$ |
| $^{140}\chi$ | $4.50\times10^{57}$ | $5.70\times10^{57}$ | $7.16\times10^{57}$ | $7.14\times10^{57}$ | $8.83\times10^{57}$ | $1.09\times10^{58}$ | $1.54\times10^{58}$ |
| $^{100}\rho$ | $1.56\times10^{55}$ | $1.80\times10^{55}$ | $2.16\times10^{55}$ | $2.40\times10^{55}$ | $2.99\times10^{55}$ | $3.59\times10^{55}$ | $4.79\times10^{55}$ |

Table 5
*Number of Nucleons of Each Isotope of Each Star when the Metallicity Equals .004*

| $M_*(M_\odot)$ | 13 | 15 | 18 | 20 | 25 | 30 | 40 |
|---|---|---|---|---|---|---|---|
| $^1H$ | $7.63\times10^{57}$ | $8.52\times10^{57}$ | $8.95\times10^{57}$ | $1.07\times10^{58}$ | $1.22\times10^{58}$ | $1.21\times10^{58}$ | $1.23\times10^{58}$ |
| $^2H$ | $1.28\times10^{43}$ | $2.38\times10^{43}$ | $5.47\times10^{43}$ | $1.05\times10^{44}$ | $3.87\times10^{41}$ | $8.30\times10^{43}$ | $3.16\times10^{41}$ |
| $^3He$ | $2.04\times10^{53}$ | $1.90\times10^{53}$ | $2.68\times10^{53}$ | $2.10\times10^{53}$ | $2.22\times10^{53}$ | $2.20\times10^{53}$ | $2.16\times10^{53}$ |
| $^4He$ | $4.84\times10^{57}$ | $5.93\times10^{57}$ | $7.26\times10^{57}$ | $8.42\times10^{57}$ | $1.02\times10^{58}$ | $9.49\times10^{57}$ | $9.73\times10^{57}$ |
| $He_{nuc}$ | $9.15\times10^{56}$ | $1.40\times10^{57}$ | $1.83\times10^{57}$ | $2.38\times10^{57}$ | $2.61\times10^{57}$ | $4.31\times10^{56}$ | $-2.3\times10^{57}$ |
| $^6Li$ | $4.12\times10^{40}$ | $7.79\times10^{40}$ | $1.80\times10^{41}$ | $3.46\times10^{41}$ | $6.22\times10^{34}$ | $2.67\times10^{41}$ | $2.56\times10^{37}$ |
| $^7Li$ | $3.77\times10^{42}$ | $8.62\times10^{43}$ | $6.04\times10^{44}$ | $1.87\times10^{45}$ | $1.22\times10^{44}$ | $9.59\times10^{44}$ | $5.64\times10^{45}$ |
| $^9Be$ | $1.28\times10^{41}$ | $1.29\times10^{41}$ | $1.89\times10^{41}$ | $6.60\times10^{41}$ | $5.85\times10^{33}$ | $4.53\times10^{41}$ | $7.19\times10^{35}$ |
| $^{10}B$ | $3.05\times10^{46}$ | $3.05\times10^{46}$ | $1.86\times10^{47}$ | $3.49\times10^{46}$ | $3.70\times10^{46}$ | $3.64\times10^{46}$ | $7.87\times10^{45}$ |
| $^{11}B$ | $1.38\times10^{47}$ | $1.38\times10^{47}$ | $8.35\times10^{47}$ | $5.51\times10^{49}$ | $1.66\times10^{47}$ | $1.64\times10^{47}$ | $3.35\times10^{46}$ |
| $^{12}C$ | $1.05\times10^{56}$ | $1.06\times10^{56}$ | $1.26\times10^{56}$ | $1.17\times10^{56}$ | $1.58\times10^{56}$ | $2.18\times10^{56}$ | $5.49\times10^{56}$ |
| $^{13}C$ | $2.25\times10^{53}$ | $2.50\times10^{53}$ | $7.22\times10^{55}$ | $3.50\times10^{53}$ | $4.58\times10^{53}$ | $4.06\times10^{53}$ | $4.41\times10^{53}$ |
| $^{14}N$ | $1.09\times10^{55}$ | $1.55\times10^{55}$ | $8.71\times10^{55}$ | $2.20\times10^{55}$ | $3.77\times10^{55}$ | $2.40\times10^{55}$ | $3.11\times10^{55}$ |
| $^{15}N$ | $8.19\times10^{51}$ | $1.04\times10^{52}$ | $6.41\times10^{55}$ | $3.33\times10^{52}$ | $1.14\times10^{53}$ | $5.96\times10^{51}$ | $6.00\times10^{51}$ |
| $^{16}O$ | $4.61\times10^{56}$ | $3.50\times10^{56}$ | $6.24\times10^{56}$ | $1.19\times10^{57}$ | $2.64\times10^{57}$ | $5.74\times10^{57}$ | $9.53\times10^{57}$ |
| $^{17}O$ | $1.05\times10^{53}$ | $1.04\times10^{53}$ | $1.33\times10^{54}$ | $1.25\times10^{53}$ | $1.34\times10^{53}$ | $1.44\times10^{53}$ | $1.90\times10^{53}$ |
| $^{18}O$ | $2.34\times10^{54}$ | $1.16\times10^{54}$ | $6.23\times10^{55}$ | $1.53\times10^{54}$ | $1.02\times10^{54}$ | $5.21\times10^{52}$ | $1.00\times10^{54}$ |
| $^{19}F$ | $2.37\times10^{51}$ | $2.55\times10^{51}$ | $1.45\times10^{52}$ | $6.42\times10^{51}$ | $1.02\times10^{53}$ | $1.39\times10^{52}$ | $1.25\times10^{51}$ |
| $^{20}Ne$ | $1.58\times10^{56}$ | $1.49\times10^{56}$ | $2.40\times10^{56}$ | $3.32\times10^{56}$ | $9.82\times10^{56}$ | $1.12\times10^{57}$ | $2.25\times10^{57}$ |
| $^{21}Ne$ | $2.22\times10^{53}$ | $1.71\times10^{53}$ | $1.20\times10^{54}$ | $3.09\times10^{53}$ | $4.59\times10^{53}$ | $8.64\times10^{53}$ | $1.40\times10^{54}$ |
| $^{22}Ne$ | $1.26\times10^{54}$ | $9.03\times10^{53}$ | $4.99\times10^{54}$ | $2.59\times10^{54}$ | $5.47\times10^{54}$ | $7.40\times10^{54}$ | $3.44\times10^{54}$ |
| $^{23}Na$ | $1.80\times10^{54}$ | $9.81\times10^{53}$ | $7.94\times10^{54}$ | $4.87\times10^{54}$ | $7.49\times10^{54}$ | $1.69\times10^{55}$ | $3.13\times10^{55}$ |
| $^{24}Mg$ | $5.25\times10^{55}$ | $8.94\times10^{55}$ | $8.30\times10^{55}$ | $1.16\times10^{56}$ | $2.79\times10^{56}$ | $2.61\times10^{56}$ | $4.54\times10^{56}$ |
| $^{25}Mg$ | $1.65\times10^{54}$ | $2.34\times10^{54}$ | $9.81\times10^{54}$ | $2.64\times10^{54}$ | $7.23\times10^{54}$ | $8.49\times10^{54}$ | $1.44\times10^{55}$ |
| $^{26}Mg$ | $1.49\times10^{54}$ | $2.31\times10^{54}$ | $7.71\times10^{54}$ | $2.25\times10^{54}$ | $8.04\times10^{54}$ | $8.47\times10^{54}$ | $1.59\times10^{55}$ |
| $^{26}Al$ | $3.21\times10^{51}$ | $5.74\times10^{51}$ | $2.71\times10^{52}$ | $1.04\times10^{52}$ | $4.42\times10^{51}$ | $1.41\times10^{52}$ | $2.67\times10^{52}$ |
| $^{27}Al$ | $2.65\times10^{54}$ | $3.96\times10^{54}$ | $7.53\times10^{54}$ | $6.14\times10^{54}$ | $1.34\times10^{55}$ | $2.02\times10^{55}$ | $3.64\times10^{55}$ |
| $^{28}Si$ | $7.32\times10^{55}$ | $1.23\times10^{56}$ | $1.13\times10^{56}$ | $1.49\times10^{56}$ | $1.43\times10^{56}$ | $4.73\times10^{56}$ | $6.26\times10^{56}$ |
| $^{29}Si$ | $6.48\times10^{53}$ | $1.34\times10^{54}$ | $3.23\times10^{54}$ | $1.52\times10^{54}$ | $2.31\times10^{54}$ | $4.04\times10^{54}$ | $5.45\times10^{54}$ |
| $^{30}Si$ | $7.86\times10^{53}$ | $1.55\times10^{54}$ | $4.66\times10^{54}$ | $1.93\times10^{54}$ | $1.89\times10^{54}$ | $5.77\times10^{54}$ | $7.79\times10^{54}$ |
| $^{31}P$ | $1.80\times10^{53}$ | $3.28\times10^{53}$ | $8.82\times10^{53}$ | $4.64\times10^{53}$ | $4.62\times10^{53}$ | $1.22\times10^{54}$ | $1.77\times10^{54}$ |
| $^{32}S$ | $3.21\times10^{55}$ | $4.12\times10^{55}$ | $4.87\times10^{55}$ | $6.17\times10^{55}$ | $4.17\times10^{55}$ | $2.28\times10^{56}$ | $2.71\times10^{56}$ |
| $^{33}S$ | $1.09\times10^{53}$ | $2.19\times10^{53}$ | $3.14\times10^{53}$ | $2.26\times10^{53}$ | $1.99\times10^{53}$ | $5.20\times10^{53}$ | $5.45\times10^{53}$ |
| $^{34}S$ | $5.11\times10^{53}$ | $1.15\times10^{54}$ | $2.85\times10^{54}$ | $1.09\times10^{54}$ | $8.65\times10^{53}$ | $2.43\times10^{54}$ | $2.81\times10^{54}$ |
| $^{36}S$ | $8.50\times10^{50}$ | $1.19\times10^{51}$ | $9.33\times10^{51}$ | $1.63\times10^{51}$ | $3.21\times10^{51}$ | $7.31\times10^{51}$ | $1.43\times10^{52}$ |
| $^{35}Cl$ | $3.19\times10^{52}$ | $6.24\times10^{52}$ | $1.31\times10^{53}$ | $6.54\times10^{52}$ | $6.35\times10^{52}$ | $1.47\times10^{53}$ | $1.87\times10^{53}$ |
| $^{37}Cl$ | $1.21\times10^{52}$ | $1.98\times10^{52}$ | $5.51\times10^{52}$ | $2.46\times10^{52}$ | $3.63\times10^{52}$ | $9.33\times10^{52}$ | $1.35\times10^{53}$ |
| $^{36}Ar$ | $5.39\times10^{54}$ | $5.16\times10^{54}$ | $8.34\times10^{54}$ | $9.57\times10^{54}$ | $5.67\times10^{54}$ | $3.70\times10^{55}$ | $4.17\times10^{55}$ |
| $^{38}Ar$ | $1.99\times10^{53}$ | $5.92\times10^{53}$ | $8.90\times10^{53}$ | $3.67\times10^{53}$ | $3.20\times10^{53}$ | $1.17\times10^{54}$ | $1.37\times10^{54}$ |
| $^{40}Ar$ | $2.75\times10^{50}$ | $4.48\times10^{50}$ | $3.05\times10^{51}$ | $3.95\times10^{50}$ | $4.28\times10^{50}$ | $1.11\times10^{51}$ | $1.38\times10^{51}$ |
| $^{39}K$ | $2.13\times10^{52}$ | $3.83\times10^{52}$ | $8.62\times10^{52}$ | $3.74\times10^{52}$ | $3.16\times10^{52}$ | $1.22\times10^{53}$ | $1.19\times10^{53}$ |
| $^{40}K$ | $9.32\times10^{48}$ | $1.09\times10^{49}$ | $8.00\times10^{49}$ | $2.22\times10^{49}$ | $4.00\times10^{49}$ | $9.47\times10^{49}$ | $8.88\times10^{49}$ |
| $^{41}K$ | $2.24\times10^{51}$ | $2.95\times10^{51}$ | $8.52\times10^{51}$ | $3.75\times10^{51}$ | $4.11\times10^{51}$ | $1.56\times10^{52}$ | $1.62\times10^{52}$ |
| $^{40}Ca$ | $4.68\times10^{54}$ | $3.69\times10^{54}$ | $7.33\times10^{54}$ | $7.79\times10^{54}$ | $4.52\times10^{54}$ | $3.10\times10^{55}$ | $3.39\times10^{55}$ |
| $^{42}Ca$ | $4.98\times10^{51}$ | $1.15\times10^{52}$ | $2.17\times10^{52}$ | $8.76\times10^{51}$ | $7.19\times10^{51}$ | $3.17\times10^{52}$ | $3.57\times10^{52}$ |
| $^{43}Ca$ | $4.08\times10^{50}$ | $4.31\times10^{50}$ | $2.31\times10^{51}$ | $6.26\times10^{50}$ | $9.52\times10^{50}$ | $1.18\times10^{51}$ | $1.71\times10^{51}$ |
| $^{44}Ca$ | $2.48\times10^{52}$ | $2.75\times10^{52}$ | $4.07\times10^{52}$ | $2.50\times10^{52}$ | $3.17\times10^{52}$ | $1.81\times10^{52}$ | $2.25\times10^{52}$ |
| $^{46}Ca$ | $8.83\times10^{49}$ | $1.41\times10^{50}$ | $8.16\times10^{50}$ | $1.32\times10^{50}$ | $1.53\times10^{50}$ | $2.36\times10^{50}$ | $3.88\times10^{50}$ |
| $^{48}Ca$ | $3.86\times10^{50}$ | $4.82\times10^{50}$ | $2.84\times10^{51}$ | $6.01\times10^{50}$ | $7.41\times10^{50}$ | $7.76\times10^{50}$ | $9.34\times10^{50}$ |
| $^{45}Sc$ | $2.59\times10^{50}$ | $2.97\times10^{50}$ | $1.62\times10^{51}$ | $4.32\times10^{50}$ | $6.35\times10^{50}$ | $1.31\times10^{51}$ | $1.96\times10^{51}$ |
| $^{46}Ti$ | $6.18\times10^{51}$ | $7.29\times10^{51}$ | $9.81\times10^{51}$ | $4.24\times10^{51}$ | $6.16\times10^{51}$ | $1.46\times10^{52}$ | $1.68\times10^{52}$ |
| $^{47}Ti$ | $6.47\times10^{51}$ | $4.07\times10^{51}$ | $5.57\times10^{51}$ | $1.59\times10^{51}$ | $7.74\times10^{51}$ | $1.75\times10^{51}$ | $2.44\times10^{51}$ |
| $^{48}Ti$ | $9.79\times10^{52}$ | $8.88\times10^{52}$ | $1.64\times10^{53}$ | $1.37\times10^{53}$ | $1.07\times10^{53}$ | $2.79\times10^{53}$ | $3.21\times10^{53}$ |
| $^{49}Ti$ | $4.18\times10^{51}$ | $3.37\times10^{51}$ | $8.68\times10^{51}$ | $6.22\times10^{51}$ | $4.90\times10^{51}$ | $1.63\times10^{52}$ | $2.06\times10^{52}$ |
| $^{50}Ti$ | $7.01\times10^{50}$ | $6.72\times10^{50}$ | $4.46\times10^{51}$ | $1.29\times10^{51}$ | $2.69\times10^{51}$ | $4.91\times10^{51}$ | $9.22\times10^{51}$ |
| $^{50}V$ | $5.52\times10^{48}$ | $5.92\times10^{48}$ | $4.86\times10^{49}$ | $1.21\times10^{49}$ | $2.37\times10^{49}$ | $6.44\times10^{49}$ | $1.16\times10^{50}$ |
| $^{51}V$ | $1.57\times10^{52}$ | $1.15\times10^{52}$ | $1.47\times10^{52}$ | $9.59\times10^{51}$ | $1.69\times10^{52}$ | $2.12\times10^{52}$ | $2.64\times10^{52}$ |
| $^{50}Cr$ | $2.10\times10^{52}$ | $3.25\times10^{52}$ | $4.00\times10^{52}$ | $3.67\times10^{52}$ | $2.96\times10^{52}$ | $1.03\times10^{53}$ | $1.45\times10^{53}$ |
| $^{52}Cr$ | $1.33\times10^{54}$ | $1.15\times10^{54}$ | $2.17\times10^{54}$ | $1.99\times10^{54}$ | $1.49\times10^{54}$ | $4.47\times10^{54}$ | $4.87\times10^{54}$ |
| $^{53}Cr$ | $8.12\times10^{52}$ | $7.07\times10^{52}$ | $1.47\times10^{53}$ | $1.25\times10^{53}$ | $8.64\times10^{52}$ | $2.91\times10^{53}$ | $3.31\times10^{53}$ |
| $^{54}Cr$ | $2.10\times10^{51}$ | $2.20\times10^{51}$ | $1.28\times10^{52}$ | $3.73\times10^{51}$ | $6.94\times10^{51}$ | $1.04\times10^{52}$ | $1.66\times10^{52}$ |
| $^{55}Mn$ | $2.40\times10^{53}$ | $2.05\times10^{53}$ | $5.61\times10^{53}$ | $3.94\times10^{53}$ | $2.80\times10^{53}$ | $9.12\times10^{53}$ | $1.00\times10^{54}$ |
| $^{54}Fe$ | $1.49\times10^{54}$ | $1.47\times10^{54}$ | $3.15\times10^{54}$ | $2.84\times10^{54}$ | $1.69\times10^{54}$ | $8.49\times10^{54}$ | $1.12\times10^{55}$ |
| $^{56}Fe$ | $8.70\times10^{55}$ | $8.74\times10^{55}$ | $1.04\times10^{56}$ | $8.86\times10^{55}$ | $8.95\times10^{55}$ | $8.94\times10^{55}$ | $8.95\times10^{55}$ |
| $^{57}Fe$ | $1.26\times10^{54}$ | $1.35\times10^{54}$ | $4.79\times10^{54}$ | $1.46\times10^{54}$ | $1.26\times10^{54}$ | $8.56\times10^{53}$ | $8.76\times10^{53}$ |
| $^{58}Fe$ | $5.71\times10^{52}$ | $5.86\times10^{52}$ | $3.19\times10^{53}$ | $1.10\times10^{53}$ | $2.22\times10^{53}$ | $3.46\times10^{53}$ | $5.20\times10^{53}$ |
| $^{59}Co$ | $2.35\times10^{53}$ | $1.57\times10^{53}$ | $3.22\times10^{53}$ | $7.44\times10^{52}$ | $2.60\times10^{53}$ | $1.53\times10^{53}$ | $2.29\times10^{53}$ |
| $^{58}Ni$ | $6.16\times10^{53}$ | $5.56\times10^{53}$ | $2.69\times10^{55}$ | $5.91\times10^{53}$ | $6.52\times10^{53}$ | $8.55\times10^{53}$ | $9.91\times10^{53}$ |
| $^{60}Ni$ | $2.16\times10^{54}$ | $2.00\times10^{54}$ | $1.64\times10^{54}$ | $1.55\times10^{54}$ | $2.16\times10^{54}$ | $3.15\times10^{53}$ | $4.55\times10^{53}$ |
| $^{61}Ni$ | $5.09\times10^{52}$ | $5.98\times10^{52}$ | $3.19\times10^{53}$ | $6.79\times10^{52}$ | $7.29\times10^{52}$ | $6.65\times10^{52}$ | $1.17\times10^{53}$ |
| $^{62}Ni$ | $4.25\times10^{52}$ | $2.87\times10^{52}$ | $3.69\times10^{54}$ | $1.09\times10^{53}$ | $1.31\times10^{53}$ | $2.10\times10^{53}$ | $3.61\times10^{53}$ |
| $^{64}Ni$ | $1.31\times10^{52}$ | $4.34\times10^{51}$ | $8.00\times10^{52}$ | $3.34\times10^{52}$ | $1.18\times10^{53}$ | $2.34\times10^{53}$ | $5.09\times10^{53}$ |
| $^{63}Cu$ | $1.26\times10^{52}$ | $6.43\times10^{51}$ | $4.34\times10^{52}$ | $1.62\times10^{52}$ | $5.11\times10^{52}$ | $6.26\times10^{52}$ | $1.25\times10^{53}$ |
| $^{65}Cu$ | $4.56\times10^{51}$ | $1.81\times10^{51}$ | $2.54\times10^{52}$ | $1.22\times10^{52}$ | $3.80\times10^{52}$ | $8.31\times10^{52}$ | $1.74\times10^{53}$ |
| $^{64}Zn$ | $1.37\times10^{53}$ | $1.39\times10^{53}$ | $2.77\times10^{52}$ | $7.23\times10^{52}$ | $1.55\times10^{53}$ | $3.57\times10^{52}$ | $6.95\times10^{52}$ |
| $^{66}Zn$ | $8.05\times10^{51}$ | $5.58\times10^{51}$ | $1.03\times10^{53}$ | $2.05\times10^{52}$ | $5.49\times10^{52}$ | $1.27\times10^{53}$ | $2.61\times10^{53}$ |
| $^{67}Zn$ | $9.45\times10^{50}$ | $3.84\times10^{50}$ | $6.92\times10^{51}$ | $2.69\times10^{51}$ | $1.03\times10^{52}$ | $2.18\times10^{52}$ | $4.92\times10^{52}$ |
| $^{68}Zn$ | $4.22\times10^{51}$ | $1.82\times10^{51}$ | $3.43\times10^{52}$ | $1.40\times10^{52}$ | $5.25\times10^{52}$ | $1.44\times10^{53}$ | $3.16\times10^{53}$ |
| $^{70}Zn$ | $1.34\times10^{50}$ | $1.07\times10^{50}$ | $1.27\times10^{51}$ | $1.11\times10^{50}$ | $1.99\times10^{50}$ | $5.49\times10^{50}$ | $6.97\times10^{50}$ |
| $^{69}Ga$ | $6.06\times10^{50}$ | $2.08\times10^{50}$ | $3.45\times10^{51}$ | $1.86\times10^{51}$ | $6.70\times10^{51}$ | $1.92\times10^{52}$ | $4.04\times10^{52}$ |
| $^{71}Ga$ | $3.65\times10^{50}$ | $2.46\times10^{50}$ | $3.46\times10^{51}$ | $1.40\times10^{51}$ | $5.65\times10^{51}$ | $1.59\times10^{52}$ | $3.50\times10^{52}$ |
| $^{140}\chi$ | $4.08\times10^{57}$ | $5.26\times10^{57}$ | $7.12\times10^{57}$ | $6.61\times10^{57}$ | $8.67\times10^{57}$ | $1.20\times10^{58}$ | $2.05\times10^{58}$ |
| $^{100}\rho$ | $6.23\times10^{55}$ | $7.19\times10^{55}$ | $8.62\times10^{55}$ | $9.58\times10^{55}$ | $1.20\times10^{56}$ | $1.44\times10^{56}$ | $1.92\times10^{56}$ |

18 BUTLER AND GEORGIEVTable 6
*Number of Nucleons of Each Isotope of Each Star when the Metallicity Equals .02*

| $M_*(M_\odot)$ | 13 | 15 | 18 | 20 | 25 | 30 | 40 |
|---|---|---|---|---|---|---|---|
| $^1H$ | $7.37\times10^{57}$ | $8.13\times10^{57}$ | $9.02\times10^{57}$ | $9.50\times10^{57}$ | $1.01\times10^{58}$ | $1.05\times10^{58}$ | $4.25\times10^{57}$ |
| $^2H$ | $9.99\times10^{42}$ | $1.22\times10^{43}$ | $4.98\times10^{41}$ | $1.44\times10^{42}$ | $1.18\times10^{42}$ | $1.29\times10^{42}$ | $9.23\times10^{40}$ |
| $^3He$ | $2.35\times10^{53}$ | $2.61\times10^{53}$ | $2.75\times10^{53}$ | $2.85\times10^{53}$ | $2.65\times10^{53}$ | $2.54\times10^{53}$ | $6.05\times10^{52}$ |
| $^4He$ | $5.15\times10^{57}$ | $6.29\times10^{57}$ | $7.32\times10^{57}$ | $8.10\times10^{57}$ | $8.67\times10^{57}$ | $1.00\times10^{58}$ | $5.64\times10^{57}$ |
| $He_{nuc}$ | $1.23\times10^{57}$ | $1.76\times10^{57}$ | $1.89\times10^{57}$ | $2.06\times10^{57}$ | $1.13\times10^{57}$ | $9.58\times10^{56}$ | $-6.4\times10^{57}$ |
| $^6Li$ | $1.58\times10^{40}$ | $3.80\times10^{40}$ | $3.33\times10^{38}$ | $1.31\times10^{35}$ | $1.14\times10^{35}$ | $7.07\times10^{34}$ | $4.16\times10^{35}$ |
| $^7Li$ | $6.74\times10^{46}$ | $8.49\times10^{44}$ | $3.35\times10^{44}$ | $5.17\times10^{45}$ | $8.30\times10^{44}$ | $5.19\times10^{44}$ | $6.85\times10^{44}$ |
| $^9Be$ | $5.52\times10^{38}$ | $1.86\times10^{39}$ | $9.06\times10^{39}$ | $5.65\times10^{34}$ | $2.67\times10^{37}$ | $5.89\times10^{34}$ | $8.42\times10^{34}$ |
| $^{10}B$ | $1.34\times10^{47}$ | $1.63\times10^{47}$ | $1.70\times10^{47}$ | $1.76\times10^{47}$ | $1.81\times10^{47}$ | $1.84\times10^{47}$ | $8.34\times10^{43}$ |
| $^{11}B$ | $5.13\times10^{47}$ | $7.21\times10^{47}$ | $7.68\times10^{47}$ | $7.83\times10^{47}$ | $8.11\times10^{47}$ | $8.29\times10^{47}$ | $3.86\times10^{43}$ |
| $^{12}C$ | $1.28\times10^{56}$ | $7.80\times10^{55}$ | $1.63\times10^{56}$ | $2.93\times10^{56}$ | $1.82\times10^{56}$ | $2.99\times10^{56}$ | $7.14\times10^{56}$ |
| $^{13}C$ | $1.20\times10^{54}$ | $1.38\times10^{54}$ | $1.65\times10^{54}$ | $1.74\times10^{54}$ | $7.98\times10^{55}$ | $2.30\times10^{54}$ | $5.16\times10^{53}$ |
| $^{14}N$ | $5.75\times10^{55}$ | $7.37\times10^{55}$ | $7.92\times10^{55}$ | $8.61\times10^{55}$ | $1.01\times10^{56}$ | $1.22\times10^{56}$ | $6.96\times10^{55}$ |
| $^{15}N$ | $4.90\times10^{52}$ | $7.15\times10^{52}$ | $1.83\times10^{52}$ | $2.68\times10^{53}$ | $5.55\times10^{55}$ | $7.85\times10^{51}$ | $6.85\times10^{51}$ |
| $^{16}O$ | $2.61\times10^{56}$ | $1.94\times10^{56}$ | $9.22\times10^{56}$ | $1.26\times10^{57}$ | $2.81\times10^{57}$ | $3.86\times10^{57}$ | $8.78\times10^{57}$ |
| $^{17}O$ | $1.04\times10^{54}$ | $9.73\times10^{53}$ | $1.05\times10^{54}$ | $1.14\times10^{54}$ | $1.61\times10^{54}$ | $2.02\times10^{54}$ | $1.16\times10^{54}$ |
| $^{18}O$ | $4.19\times10^{54}$ | $3.04\times10^{54}$ | $1.40\times10^{55}$ | $6.25\times10^{54}$ | $1.01\times10^{56}$ | $7.40\times10^{54}$ | $1.47\times10^{55}$ |
| $^{19}F$ | $1.70\times10^{52}$ | $1.96\times10^{52}$ | $5.35\times10^{51}$ | $7.25\times10^{52}$ | $1.43\times10^{53}$ | $9.35\times10^{51}$ | $6.30\times10^{51}$ |
| $^{20}Ne$ | $4.16\times10^{55}$ | $4.06\times10^{55}$ | $1.78\times10^{56}$ | $4.72\times10^{56}$ | $1.02\times10^{57}$ | $1.12\times10^{57}$ | $2.65\times10^{57}$ |
| $^{21}Ne$ | $2.69\times10^{53}$ | $1.56\times10^{53}$ | $2.42\times10^{53}$ | $2.17\times10^{54}$ | $1.90\times10^{54}$ | $3.56\times10^{54}$ | $5.95\times10^{54}$ |
| $^{22}Ne$ | $5.41\times10^{54}$ | $2.06\times10^{54}$ | $7.94\times10^{54}$ | $1.08\times10^{55}$ | $2.01\times10^{55}$ | $2.23\times10^{55}$ | $1.31\times10^{55}$ |
| $^{23}Na$ | $1.11\times10^{54}$ | $1.27\times10^{54}$ | $3.51\times10^{54}$ | $2.01\times10^{55}$ | $2.23\times10^{55}$ | $4.16\times10^{55}$ | $9.29\times10^{55}$ |
| $^{24}Mg$ | $3.02\times10^{55}$ | $4.54\times10^{55}$ | $1.23\times10^{56}$ | $8.58\times10^{55}$ | $2.61\times10^{56}$ | $2.25\times10^{56}$ | $3.71\times10^{56}$ |
| $^{25}Mg$ | $3.07\times10^{54}$ | $1.75\times10^{54}$ | $8.48\times10^{54}$ | $1.72\times10^{55}$ | $3.75\times10^{55}$ | $3.74\times10^{55}$ | $8.72\times10^{55}$ |
| $^{26}Mg$ | $2.59\times10^{54}$ | $2.06\times10^{54}$ | $7.02\times10^{54}$ | $1.06\times10^{55}$ | $3.26\times10^{55}$ | $3.34\times10^{55}$ | $8.78\times10^{55}$ |
| $^{26}Al$ | $2.55\times10^{52}$ | $9.14\times10^{51}$ | $4.42\times10^{52}$ | $1.80\times10^{52}$ | $1.04\times10^{53}$ | $4.67\times10^{52}$ | $7.95\times10^{52}$ |
| $^{27}Al$ | $1.80\times10^{54}$ | $2.92\times10^{54}$ | $1.20\times10^{55}$ | $1.19\times10^{55}$ | $3.23\times10^{55}$ | $4.08\times10^{55}$ | $9.94\times10^{55}$ |
| $^{28}Si$ | $8.96\times10^{55}$ | $1.00\times10^{56}$ | $1.21\times10^{56}$ | $7.57\times10^{55}$ | $1.53\times10^{56}$ | $2.87\times10^{56}$ | $2.89\times10^{56}$ |
| $^{29}Si$ | $1.78\times10^{54}$ | $2.64\times10^{54}$ | $8.34\times10^{54}$ | $2.44\times10^{54}$ | $8.46\times10^{54}$ | $8.84\times10^{54}$ | $1.20\times10^{55}$ |
| $^{30}Si$ | $1.86\times10^{54}$ | $3.29\times10^{54}$ | $8.17\times10^{54}$ | $2.92\times10^{54}$ | $7.41\times10^{54}$ | $1.27\times10^{55}$ | $1.17\times10^{55}$ |
| $^{31}P$ | $4.49\times10^{53}$ | $8.46\times10^{53}$ | $2.08\times10^{54}$ | $7.88\times10^{53}$ | $1.80\times10^{54}$ | $3.07\times10^{54}$ | $4.22\times10^{54}$ |
| $^{32}S$ | $4.47\times10^{55}$ | $4.16\times10^{55}$ | $4.42\times10^{55}$ | $3.37\times10^{55}$ | $5.98\times10^{55}$ | $1.29\times10^{56}$ | $1.31\times10^{56}$ |
| $^{33}S$ | $2.40\times10^{53}$ | $2.91\times10^{53}$ | $4.20\times10^{53}$ | $2.71\times10^{53}$ | $3.89\times10^{53}$ | $5.76\times10^{53}$ | $5.76\times10^{53}$ |
| $^{34}S$ | $1.96\times10^{54}$ | $1.83\times10^{54}$ | $2.62\times10^{54}$ | $2.04\times10^{54}$ | $2.71\times10^{54}$ | $4.50\times10^{54}$ | $4.10\times10^{54}$ |
| $^{36}S$ | $6.43\times10^{51}$ | $3.04\times10^{51}$ | $1.44\times10^{52}$ | $1.10\times10^{52}$ | $2.90\times10^{52}$ | $6.10\times10^{52}$ | $1.13\times10^{53}$ |
| $^{35}Cl$ | $1.38\times10^{53}$ | $1.64\times10^{53}$ | $2.10\times10^{53}$ | $1.50\times10^{53}$ | $1.94\times10^{53}$ | $2.79\times10^{53}$ | $3.31\times10^{53}$ |
| $^{37}Cl$ | $3.63\times10^{52}$ | $2.98\times10^{52}$ | $6.77\times10^{52}$ | $1.01\times10^{53}$ | $2.04\times10^{53}$ | $3.15\times10^{53}$ | $6.76\times10^{53}$ |
| $^{36}Ar$ | $7.53\times10^{54}$ | $5.87\times10^{54}$ | $6.56\times10^{54}$ | $5.58\times10^{54}$ | $9.49\times10^{54}$ | $2.17\times10^{55}$ | $2.17\times10^{55}$ |
| $^{38}Ar$ | $8.22\times10^{53}$ | $7.80\times10^{53}$ | $9.38\times10^{53}$ | $8.07\times10^{53}$ | $1.08\times10^{54}$ | $1.96\times10^{54}$ | $1.98\times10^{54}$ |
| $^{40}Ar$ | $1.07\times10^{51}$ | $1.26\times10^{51}$ | $1.56\times10^{51}$ | $1.69\times10^{51}$ | $3.21\times10^{51}$ | $8.88\times10^{51}$ | $6.79\times10^{51}$ |
| $^{39}K$ | $6.01\times10^{52}$ | $8.38\times10^{52}$ | $9.19\times10^{52}$ | $8.85\times10^{52}$ | $1.06\times10^{53}$ | $1.55\times10^{53}$ | $1.37\times10^{53}$ |
| $^{40}K$ | $1.53\times10^{50}$ | $7.70\times10^{49}$ | $1.65\times10^{50}$ | $1.11\times10^{50}$ | $2.40\times10^{50}$ | $3.52\times10^{50}$ | $4.85\times10^{50}$ |
| $^{41}K$ | $5.61\times10^{51}$ | $7.01\times10^{51}$ | $9.27\times10^{51}$ | $1.08\times10^{52}$ | $1.66\times10^{52}$ | $2.36\times10^{52}$ | $5.41\times10^{52}$ |
| $^{40}Ca$ | $5.89\times10^{54}$ | $4.80\times10^{54}$ | $5.44\times10^{54}$ | $4.48\times10^{54}$ | $7.85\times10^{54}$ | $1.88\times10^{55}$ | $1.87\times10^{55}$ |
| $^{42}Ca$ | $1.66\times10^{52}$ | $2.10\times10^{52}$ | $2.48\times10^{52}$ | $2.08\times10^{52}$ | $2.96\times10^{52}$ | $4.97\times10^{52}$ | $5.37\times10^{52}$ |
| $^{43}Ca$ | $1.63\times10^{51}$ | $1.80\times10^{51}$ | $2.54\times10^{51}$ | $3.02\times10^{51}$ | $4.10\times10^{51}$ | $3.56\times10^{51}$ | $5.91\times10^{51}$ |
| $^{44}Ca$ | $4.20\times10^{52}$ | $4.25\times10^{52}$ | $6.53\times10^{52}$ | $6.94\times10^{52}$ | $6.12\times10^{52}$ | $4.84\times10^{52}$ | $5.59\times10^{52}$ |
| $^{46}Ca$ | $2.54\times10^{50}$ | $3.99\times10^{50}$ | $9.38\times10^{50}$ | $7.37\times10^{50}$ | $1.41\times10^{51}$ | $1.41\times10^{51}$ | $3.43\times10^{51}$ |
| $^{48}Ca$ | $1.89\times10^{51}$ | $2.23\times10^{51}$ | $3.27\times10^{51}$ | $2.81\times10^{51}$ | $3.27\times10^{51}$ | $1.70\times10^{52}$ | $2.91\times10^{51}$ |
| $^{45}Sc$ | $8.73\times10^{50}$ | $1.11\times10^{51}$ | $1.51\times10^{51}$ | $1.56\times10^{51}$ | $3.35\times10^{51}$ | $3.08\times10^{51}$ | $7.37\times10^{51}$ |
| $^{46}Ti$ | $7.20\times10^{51}$ | $1.03\times10^{52}$ | $1.14\times10^{52}$ | $1.02\times10^{52}$ | $1.35\times10^{52}$ | $2.23\times10^{52}$ | $2.53\times10^{52}$ |
| $^{47}Ti$ | $5.28\times10^{51}$ | $4.12\times10^{51}$ | $6.40\times10^{51}$ | $7.85\times10^{51}$ | $6.70\times10^{51}$ | $1.06\times10^{52}$ | $8.08\times10^{51}$ |
| $^{48}Ti$ | $9.74\times10^{52}$ | $1.28\times10^{53}$ | $1.58\times10^{53}$ | $1.38\times10^{53}$ | $1.83\times10^{53}$ | $3.03\times10^{53}$ | $2.92\times10^{53}$ |
| $^{49}Ti$ | $5.68\times10^{51}$ | $6.07\times10^{51}$ | $7.32\times10^{51}$ | $7.63\times10^{51}$ | $1.32\times10^{52}$ | $2.30\times10^{52}$ | $2.90\times10^{52}$ |
| $^{50}Ti$ | $2.79\times10^{51}$ | $2.89\times10^{51}$ | $4.68\times10^{51}$ | $6.97\times10^{51}$ | $1.47\times10^{52}$ | $1.19\times10^{52}$ | $4.56\times10^{52}$ |
| $^{50}V$ | $2.43\times10^{49}$ | $3.19\times10^{49}$ | $7.91\times10^{49}$ | $4.66\times10^{49}$ | $1.15\times10^{50}$ | $2.98\times10^{50}$ | $2.24\times10^{50}$ |
| $^{51}V$ | $1.41\times10^{52}$ | $1.19\times10^{52}$ | $1.37\times10^{52}$ | $1.59\times10^{52}$ | $1.72\times10^{52}$ | $2.99\times10^{52}$ | $2.68\times10^{52}$ |
| $^{50}Cr$ | $2.67\times10^{52}$ | $4.73\times10^{52}$ | $4.41\times10^{52}$ | $3.71\times10^{52}$ | $6.25\times10^{52}$ | $9.23\times10^{52}$ | $1.10\times10^{53}$ |
| $^{52}Cr$ | $7.76\times10^{53}$ | $1.52\times10^{54}$ | $1.53\times10^{54}$ | $1.28\times10^{54}$ | $2.44\times10^{54}$ | $4.50\times10^{54}$ | $4.53\times10^{54}$ |
| $^{53}Cr$ | $7.47\times10^{52}$ | $1.11\times10^{53}$ | $1.15\times10^{53}$ | $9.92\times10^{52}$ | $1.69\times10^{53}$ | $3.28\times10^{53}$ | $2.98\times10^{53}$ |
| $^{54}Cr$ | $7.85\times10^{51}$ | $8.14\times10^{51}$ | $1.82\times10^{52}$ | $1.99\times10^{52}$ | $3.32\times10^{52}$ | $6.13\times10^{52}$ | $6.78\times10^{52}$ |
| $^{55}Mn$ | $3.37\times10^{53}$ | $4.55\times10^{53}$ | $4.85\times10^{53}$ | $4.31\times10^{53}$ | $6.59\times10^{53}$ | $1.25\times10^{54}$ | $9.92\times10^{53}$ |
| $^{54}Fe$ | $2.30\times10^{54}$ | $2.95\times10^{54}$ | $2.98\times10^{54}$ | $2.66\times10^{54}$ | $4.60\times10^{54}$ | $8.23\times10^{54}$ | $7.69\times10^{54}$ |
| $^{56}Fe$ | $9.97\times10^{55}$ | $1.02\times10^{56}$ | $1.04\times10^{56}$ | $1.06\times10^{56}$ | $1.08\times10^{56}$ | $1.10\times10^{56}$ | $9.68\times10^{55}$ |
| $^{57}Fe$ | $2.66\times10^{54}$ | $2.38\times10^{54}$ | $3.20\times10^{54}$ | $2.80\times10^{54}$ | $2.24\times10^{54}$ | $3.38\times10^{54}$ | $1.17\times10^{54}$ |
| $^{58}Fe$ | $1.45\times10^{53}$ | $1.32\times10^{53}$ | $7.02\times10^{53}$ | $5.94\times10^{53}$ | $1.00\times10^{54}$ | $2.86\times10^{54}$ | $1.77\times10^{54}$ |
| $^{59}Co$ | $1.69\times10^{53}$ | $1.08\times10^{53}$ | $2.18\times10^{53}$ | $3.50\times10^{53}$ | $4.66\times10^{53}$ | $6.46\times10^{53}$ | $8.31\times10^{53}$ |
| $^{58}Ni$ | $2.67\times10^{54}$ | $1.37\times10^{54}$ | $3.23\times10^{54}$ | $2.20\times10^{54}$ | $1.87\times10^{54}$ | $1.86\times10^{54}$ | $1.06\times10^{54}$ |
| $^{60}Ni$ | $2.55\times10^{54}$ | $2.24\times10^{54}$ | $2.54\times10^{54}$ | $2.98\times10^{54}$ | $2.17\times10^{54}$ | $7.33\times10^{53}$ | $1.44\times10^{54}$ |
| $^{61}Ni$ | $9.83\times10^{52}$ | $1.49\times10^{53}$ | $1.25\times10^{53}$ | $1.78\times10^{53}$ | $2.59\times10^{53}$ | $1.32\times10^{53}$ | $4.68\times10^{53}$ |
| $^{62}Ni$ | $2.72\times10^{53}$ | $1.96\times10^{53}$ | $5.56\times10^{53}$ | $4.62\times10^{53}$ | $7.92\times10^{53}$ | $2.66\times10^{53}$ | $1.40\times10^{54}$ |
| $^{64}Ni$ | $4.01\times10^{52}$ | $2.06\times10^{52}$ | $3.58\times10^{52}$ | $2.24\times10^{53}$ | $7.38\times10^{53}$ | $8.52\times10^{52}$ | $2.66\times10^{54}$ |
| $^{63}Cu$ | $2.19\times10^{52}$ | $1.56\times10^{52}$ | $2.28\times10^{52}$ | $1.53\times10^{53}$ | $3.52\times10^{53}$ | $1.03\times10^{53}$ | $8.20\times10^{53}$ |
| $^{65}Cu$ | $1.23\times10^{52}$ | $8.60\times10^{51}$ | $9.13\times10^{51}$ | $5.27\times10^{52}$ | $1.75\times10^{53}$ | $1.56\times10^{52}$ | $7.77\times10^{53}$ |
| $^{64}Zn$ | $1.16\times10^{53}$ | $8.11\times10^{52}$ | $6.88\times10^{52}$ | $1.44\times10^{53}$ | $6.43\times10^{52}$ | $2.24\times10^{52}$ | $2.01\times10^{53}$ |
| $^{66}Zn$ | $2.08\times10^{52}$ | $2.13\times10^{52}$ | $2.60\times10^{52}$ | $7.50\times10^{52}$ | $2.86\times10^{53}$ | $1.84\times10^{52}$ | $1.28\times10^{54}$ |
| $^{67}Zn$ | $3.45\times10^{51}$ | $2.13\times10^{51}$ | $2.43\times10^{51}$ | $1.55\times10^{52}$ | $6.30\times10^{52}$ | $3.26\times10^{51}$ | $2.97\times10^{53}$ |
| $^{68}Zn$ | $1.58\times10^{52}$ | $8.97\times10^{51}$ | $1.40\times10^{52}$ | $7.22\times10^{52}$ | $3.76\times10^{53}$ | $1.52\times10^{52}$ | $1.86\times10^{54}$ |
| $^{70}Zn$ | $2.19\times10^{51}$ | $6.70\times10^{50}$ | $6.14\times10^{50}$ | $1.94\times10^{51}$ | $1.01\times10^{52}$ | $4.42\times10^{50}$ | $3.17\times10^{52}$ |
| $^{69}Ga$ | $2.00\times10^{51}$ | $1.01\times10^{51}$ | $1.09\times10^{51}$ | $7.98\times10^{51}$ | $3.28\times10^{52}$ | $1.65\times10^{51}$ | $2.10\times10^{53}$ |
| $^{71}Ga$ | $1.77\times10^{51}$ | $9.53\times10^{50}$ | $1.27\times10^{51}$ | $6.16\times10^{51}$ | $3.34\times10^{52}$ | $1.32\times10^{51}$ | $1.58\times10^{53}$ |
| $^{140}\chi$ | $4.25\times10^{57}$ | $5.55\times10^{57}$ | $6.92\times10^{57}$ | $8.01\times10^{57}$ | $1.16\times10^{58}$ | $1.55\times10^{58}$ | $3.38\times10^{58}$ |
| $^{100}\rho$ | $3.11\times10^{56}$ | $3.59\times10^{56}$ | $4.31\times10^{56}$ | $4.79\times10^{56}$ | $5.99\times10^{56}$ | $7.19\times10^{56}$ | $9.58\times10^{56}$ |

Table 7
*Initial Total Number of Nucleons of Each Star*

| $M_*(M_\odot)$ | 13 | 15 | 18 | 20 | 25 | 30 | 40 |
|---|---|---|---|---|---|---|---|
| $\beta_i$ | $1.56\times10^{58}$ | $1.80\times10^{58}$ | $2.16\times10^{58}$ | $2.40\times10^{58}$ | $2.99\times10^{58}$ | $3.59\times10^{58}$ | $4.79\times10^{58}$ |

Table 8
$\varrho_{sum}$ *of Each Star when only Metallicity Varies, Subscripts Next to* $\varrho_{sum}$ *Indicate the Metallicity of Stars in the Row*

| $M_*(M_\odot)$ | 13 | 15 | 18 | 20 | 25 | 30 | 40 |
|---|---|---|---|---|---|---|---|
| $\varrho_{sum.0}$ | $2.53\times10^{58}$ | $3.95\times10^{58}$ | $6.14\times10^{58}$ | $8.53\times10^{58}$ | $1.09\times10^{59}$ | $1.53\times10^{59}$ | $2.56\times10^{59}$ |
| $\varrho_{sum.001}$ | $3.80\times10^{58}$ | $3.82\times10^{58}$ | $6.59\times10^{58}$ | $1.30\times10^{59}$ | $2.22\times10^{59}$ | $3.29\times10^{59}$ | $6.01\times10^{59}$ |
| $\varrho_{sum.004}$ | $4.23\times10^{58}$ | $1.24\times10^{59}$ | $2.72\times10^{59}$ | $1.24\times10^{59}$ | $2.39\times10^{59}$ | $5.57\times10^{59}$ | $1.41\times10^{60}$ |
| $\varrho_{sum.02}$ | $3.95\times10^{58}$ | $1.33\times10^{59}$ | $2.11\times10^{59}$ | $2.86\times10^{59}$ | $6.06\times10^{59}$ | $9.61\times10^{59}$ | $3.17\times10^{60}$ |

Table 9
*Progress of Nucleosynthesis within Each Star when the Metallicity Varies, Subscripts Next to P Indicate the Metallicity of Stars in the Row*

| $M_*(M_\odot)$ | 13 | 15 | 18 | 20 | 25 | 30 | 40 |
|---|---|---|---|---|---|---|---|
| $P_0$ | 1.63 | 2.20 | 2.85 | 3.56 | 3.64 | 4.26 | 5.34 |
| $P_{.001}$ | 2.44 | 2.12 | 3.06 | 5.41 | 7.42 | 9.16 | 12.49 |
| $P_{.004}$ | 2.72 | 6.88 | 12.64 | 5.17 | 7.98 | 1.550 | 29.40 |
| $P_{.02}$ | 2.54 | 7.40 | 9.78 | 11.93 | 20.24 | 26.75 | 66.15 |

Table 10
*$\varrho_{sum}$ of Each Star when Explosion Energy Varies, Subscripts Next to $\varrho_{sum}$ Indicate the Metallicity of Stars in the Row*

| $M_*$ ($M_\odot$) | 20 | 25 | 30 | 40 |
|---|---|---|---|---|
| E ($10^{51}ergs$) | 10 | 10 | 20 | 30 |
| $\varrho_{sum0}$ | $7.75\times10^{58}$ | $8.28\times10^{58}$ | $1.26\times10^{59}$ | $2.01\times10^{59}$ |
| $\varrho_{sum.001}$ | $5.44\times10^{58}$ | $1.03\times10^{59}$ | $2.22\times10^{59}$ | $3.87\times10^{59}$ |
| $\varrho_{sum.004}$ | $1.35\times10^{59}$ | $2.05\times10^{59}$ | $7.31\times10^{59}$ | $1.57\times10^{60}$ |
| $\varrho_{sum.02}$ | $2.78\times10^{59}$ | $6.05\times10^{59}$ | $9.51\times10^{59}$ | $3.20\times10^{60}$ |

Table 11
*Progress of Nucleosynthesis in Stars when Metallicity and Explosion Energy Varies, Subscripts Next to P Indicate the Metallicity of Stars in the Row*

| $M_*(M_\odot)$ | 20 | 25 | 30 | 40 |
|---|---|---|---|---|
| E ($10^{51}ergs$) | 10 | 10 | 20 | 30 |
| $P_0$ | 3.24 | 2.77 | 3.50 | 4.20 |
| $P_{.001}$ | 2.27 | 3.43 | 6.18 | 8.07 |
| $P_{.004}$ | 5.63 | 6.83 | 20.35 | 32.80 |
| $P_{.02}$ | 11.603 | 20.188 | 26.478 | 66.693 |